\documentclass[twocolumn]{aastex6}

\usepackage{graphicx}
\graphicspath{{./}{figures/}}

\received{December 14, 2018}
\revised{March 6, 2019}
\accepted{March 7, 2019}

\shorttitle{Snow Lines and Different $\alpha$ Profiles}
\shortauthors{Kalyaan \& Desch}

\begin{document}

\title{Effect of Different Angular Momentum Transport mechanisms on the Distribution of Water in Protoplanetary Disks}

\author{Anusha Kalyaan} 
\affil{School of Earth and Space Exploration, Arizona State University, PO Box 871404, Tempe, AZ 85287-1404}

\and

\author{Steven J.\ Desch}
\affil{School of Earth and Space Exploration, Arizona State University, PO Box 871404, Tempe, AZ 85287-1404}





\begin{abstract}

The snow line in a protoplanetary disk demarcates regions with H$_2$O ice from regions with H$_2$O vapor. Where a planet forms relative to this location determines how much water and other volatiles it forms with. Giant planet formation may be triggered at the water snow line if vapor diffuses outward and is cold-trapped beyond the snow line faster than icy particles can drift inward. In this study we investigate the distribution of water across the snow line, considering three different radial profiles of the turbulence parameter $\alpha(r)$, corresponding to three different angular momentum transport mechanisms. We consider the radial transport of water vapor and icy particles by diffusion, advection, and drift. We show that even for similar values of $\alpha$, the gradient of $\alpha$(r) across the snow line significantly changes the snow line location, the sharpness of the volatile gradient across the snow line, and the final water/rock ratio in planetary bodies. A profile of radially decreasing $\alpha$, consistent with transport by hydrodynamic instabilities plus magnetic disk winds, appears consistent with the distribution of water in the solar nebula, with monotonically-increasing radial water content and a diverse population of asteroids with different water content. We argue that $\Sigma(r)$ and water abundance $N_{\rm H_2O}(r)/N_{\rm H_2}(r)$ are likely diagnostic of $\alpha(r)$ and thus the mechanism for angular momentum transport in inner disks.
\end{abstract}


\section{Introduction} \label{sec:intro}
A snow line is the boundary in a protoplanetary disk between the region
near the star where a condensible volatile, especially water, is present 
as vapor, and far from the star where it is present as a solid.
The location of a snow line depends on the pressure-temperature conditions
in the disk (Hayashi 1981; Stevenson \& Lunine 1988). Icy solids drift inwards into the inner disk through the snow line region as they lose angular momentum by moving against the pressure-supported gas disk (Weidenschilling 1977). Ice on solids sublimates into vapor as it approaches the snow line. This vapor is able to diffuse through the gas both inwards towards the star and outwards back through the snow line (Stevenson \& Lunine 1988). A bidirectional flow of water is thus established across the snow line region. The location of the snow line in the disk is straightforwardly set by the 
pressure and temperature of the disk, and is typically where the midplane 
temperature is about 160 K to 180 K (Lodders 2003).
In contrast, the distributions of water and volatiles across the snow line
region---whether the abundance of water ice is enhanced or depleted beyond 
the snow line, or whether the water vapor abundance inside the snow line is 
enhanced or depleted---depend subtly on the relative rates of different
radial transport processes. 

The mechanics of radial transport affects not just the distribution of water,
but other volatiles as well. 
The chemical inventories of planetesimals (asteroids) and planets forming in 
a disk will depend on the radial distributions of these species in the disk.
Some major condensible species, e.g. CO, have their own snow lines 
(\"{O}berg et al.\ 2011). 
Other volatiles, e.g., NH$_3$, are trace species, but are expected to condense 
with water (Dodson-Robinson et al.\ 2009).
The chemical equilibria of these and other species are affected by the 
abundances of volatiles in the disk at different radii $r$ (Cuzzi \&
Zahnle 2004; Najita et al.\ 2013).
Modeling the radial distribution of all volatile species therefore depends on 
understanding how volatile distribution at the water snow line operates. 

Besides affecting the distribution of chemical species in the disk, snow lines
also can directly affect the growth of planets. 
Water ice also can enhance coagulation rates of icy particles over those of bare
silicate particles because of ice's higher sticking coefficient (Gunlach \&
Blum 2015).
An enhancement in solid mass density beyond the snow line is also possible
due to the cold-trapping of vapor diffused across the snow line 
(Stevenson \& Lunine 1988; Ros \& Johansen 2013).
This can directly enhance the coagulation rate as well, but may also increase
the solids-to-gas ratio above the critical threshold for triggering 
planetesimal growth via the streaming instability (Johansen et al.\ 2007).
The increase in solids-to-gas ratio also can lower the ionization of the gas
beyond the snow line. 
If the disk is evolving by magnetorotational instability, this would lead to a
local decrease in the angular momentum transport, and a local build up of gas. 
This in turn can lead to a localized pressure maximum in which particles can
concentrate, further enhancing planet growth (Kretke \& Lin 2008).
These factors may have led to rapid formation of Jupiter at the snow line
in the solar nebula.

\begin{figure*}
\epsscale{0.78}
\plotone{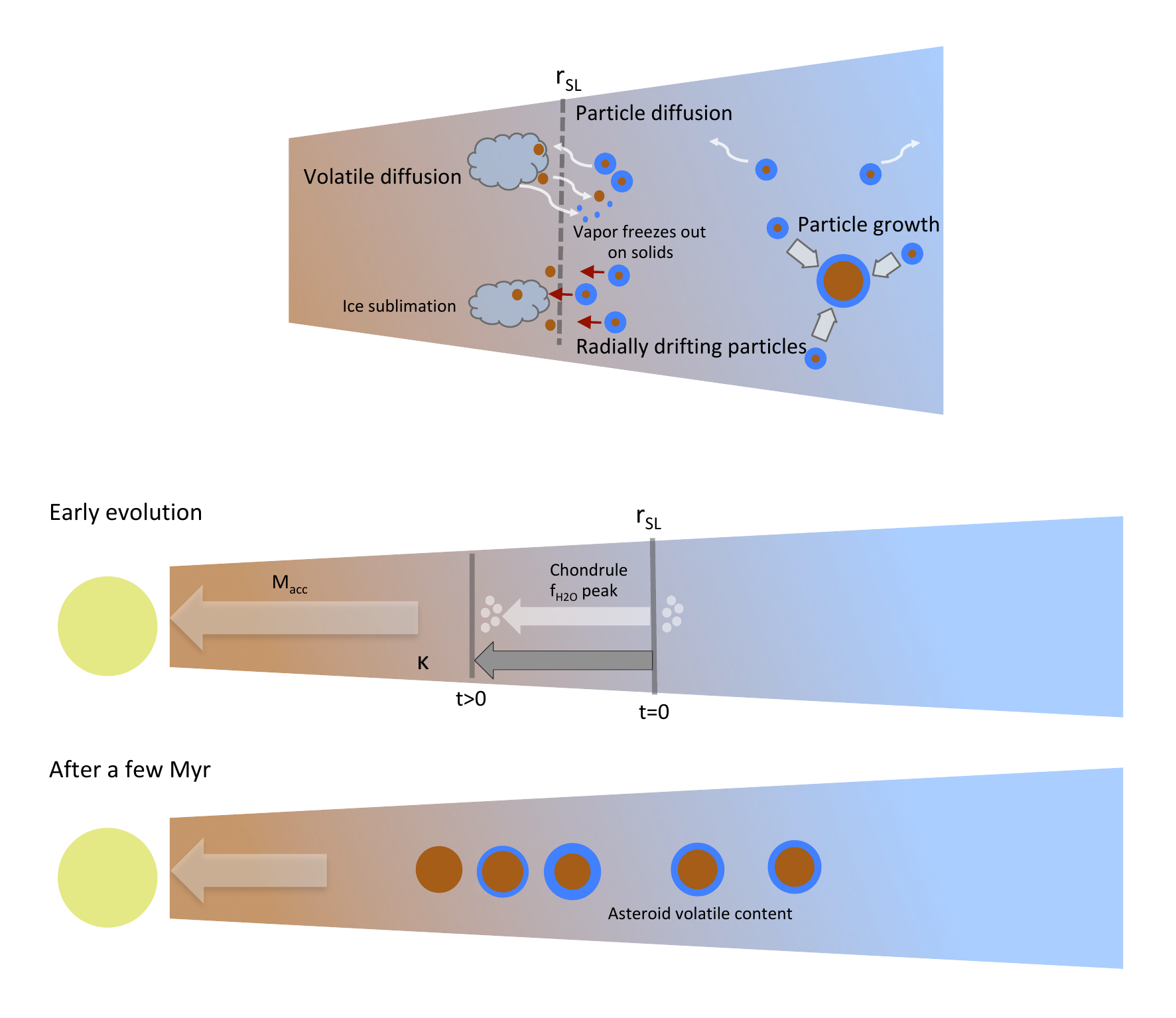}
\caption{Schematic figures (a, b, c) that show the various processes of radial transport of volatiles across the snow line, and the implications of the redistribution of volatiles over disk evolution. a) A schematic that shows the various radial transport processes that move volatiles bidirectionally through the snow line. All these processes can be sequentially contextualized as follows: 1) particles from the outer disk face a headwind from the pressure-supported gas and therefore spiral inwards; 2) small icy particles well-coupled to the gas also diffuse inwards 2) ice on these particles sublimate on reaching warmer regions of the inner disk; 3) some of this vapor diffuses back through the snow line to refreeze onto solids; 4) ice-bereft particles also diffuse back through, and may gain some of their icy mantles; 5) icy chondrules continue to diffuse both inwards back through the snow line as well as into the outer disk; 6) with time, the particles grow and/or are accreted into asteroids. b) In the early stages of disk evolution, viscous dissipation contributes significantly to the thermal structure of the nebula. Mass accretion rates are initially high. Eventually they drop down and so does the temperature of the inner disk, causing the snow line to move inwards with time. A peak in the ice abundance of chondrules forms just beyond the snowline; as the snow line moves inward, this peak follows with time. c) With time, the collective signature of the redistribution of volatiles through the above processes manifest as the bulk abundance of water available in different bodies at different heliocentric distances. See text for detailed discussion.}
\label{bigpicture}
\end{figure*}

As depicted in Figure \ref{bigpicture}, the location of the snow line and the radial 
distribution of water and volatiles depend on the thermal structure of 
the disk and the relative rates of several transport processes, including
advection and diffusion of vapor and particles, and radial drift of particles
by aerodynamic drag (e.g., Cuzzi \& Zahnle 2004; Ciesla \& Cuzzi 2006).
All of these processes are strongly affected by the angular momentum transport
in the disk, parameterized by the dimensionless parameter $\alpha$.
Not only the strength of the turbulence (the magnitude of $\alpha$), but the
spatial structure of $\alpha$ (how it varies in the disk with $r$, the distance 
from the star) can affect these processes.
The goal of this paper is to quantify how the strength and spatial structure
of turbulent viscosity, parameterized by $\alpha(r)$, affect the radial
distribution of water across a protoplanetary disk.
This will improve first-principles models, allowing them to predict how much water 
an exoplanet may have accreted, based on other observable data. 

\subsection{Observations of the Water Snow Line Region}
Observational constraints greatly assist these modeling efforts but by themselves do
not allow firm predictions of exoplanet water contents. 
Constraints on the radial distribution of water in disks most commonly come from 
studies of the water content of solar system bodies, or from infrared or millimeter
observations of water vapor.

Within the solar system, a clear gradient in water content is observed. Of the terrestrial planets, the bulk Earth (crust + mantle + core) likely accreted a few earth oceans in the mantle and therefore up to 0.1-0.2 \% by wt (Mottl et al. 2007; Wu et al 2018). Mars may have accreted $\sim 1$wt\% water (Wanke and Dreibus 1994). S-type asteroids present largely inside 2.7 AU (Gradie \& Tedesco 1982; De Meo and Carry 2014)
are associated with ordinary chondrites which accreted with $\sim 1$ wt\% water (Hutchison et al.\ 1987; Alexander et al.\ 1989,2013),
whereas C-type asteroids beyond 2.7 AU are associated with carbonaceous chondrites with up to 13wt\% H$_2$O (as hydrated silicates; Alexander et al. 2013).
The icy satellites of Jupiter and the other planets, Pluto as well as the Kuiper Belt Objects, are roughly 50wt\% ice (Brown 2012). Therefore, radially outward from the sun, there is an indication of a sharp increase in the bulk water abundance existed towards the outer disk, potentially in the asteroid belt or closer. However, it is not clear how much this distribution was affected by the presence of Jupiter
(e.g., Morbidelli et al.\ 2016), or by the particular transport mechanisms acting in our
protoplanetary disk that might not act in other disks.

Infrared observations of existing protoplanetary disks provide useful information about
the abundance of warm water vapor inside the snow line, but the snow line itself is not
resolved by such observations, as its close proximity ($< 5$ AU) to the central star means 
it subtends only tens of milliarcseconds at typical disk distances (Pontoppiddan et al.\ 2014;
Notsu et al.\ 2016). 
Instead, high-resolution spectroscopy in combination with models is used to infer the
distribution of water. 
An 'approximate' radial location of the snow line has been inferred from high resolution spectroscopy, using the inferred temperature associated with the mid-infrared lines of water vapor present in the disk atmosphere (Meijerink et al.\ 2009, Zhang et al.\ 2013). Meijerink et al.\ inferred that water vapor remained high in the inner disk up to a few AU from the star, and then rapidly decreased in concentration in the disk atmosphere. They argued that the vapor may have frozen out onto dust grains that sedimented toward the
disk midplane. Zhang et al.\ (2013) performed the same analysis for TW Hya and found a dramatic drop in vapor abundance at around 4 AU. Zhang et al.\ (2015) infer multiple snow lines due to different volatiles at different gaps in the substructure seen in HL Tau (ALMA Partnership 2015) with a chemical model, arguing that gaps in the substructure resulted from enhanced pebble accretion at these condensation fronts. This effort is however based on a number of assumptions of disk chemistry and does not directly access the water snow line. Notsu et al.\ (2016, 2017, 2018) have discovered a new method implementing high-dispersion spectroscopy to place the water snow line by the selection of specific emission lines of water vapor (with low Einstein A$_{ul}$ and high upper-state energies) that are likely to originate from the innermost warm disk. They argue that this is within ALMA's current capabilities. 

Finally, serendipitous observations have constrained the instantaneous position of a snow line
of an FU Orionis star in the midst of an outburst (Cieza et al.\ 2016). In this case, the drastic change in the thermal structure of the disk following the outburst may have moved the water snow line out to a distance of 40 AU from the star, far enough for the snow line to be spatially resolved. 

The rapid improvement in observational techniques makes it likely that the location of the snow
line will be constrained around many disks in the foreseeable future; but additional observations 
would be needed to constrain the abundance of water ice or water vapor on either side of the snow 
line.
Substantial modeling is still required to build predictive models of exoplanet water content. 

\subsection{Models of the Water Snow Line Region}
Several models (eg. Ciesla \& Cuzzi 2006; Garaud \& Lin 2007; Dodson-Robinson et al.\ 2009; Min et al.\ 2011; Desch et al.\ 2018) of the snow line calculate the thermal structure of the solar nebula in order to determine the radial location where water vapor finds the right range of temperatures and pressures to condense onto small solids. These models differ in assumptions of key parameters that affect the mid-plane temperature, including the choice of opacity of the disk material $\kappa$  (and its variation with $r$, if included) and the mass accretion rates. Min et al.\ (2011) who performed full 3-D radiative transfer simulations summarized the effects of these two properties in Table 2 of their paper, and compared the results of simulations of other works. In their canonical run, r$_{\rm ice}$ (i.e., the radius where fractional abundance of ice reaches 50\% at the midplane) varies from 16.1 AU to 0.7 AU, for $\dot{M_{\odot}}$ = 10$^{-6}$ to 10$^{-9}$ M$_{\odot}$/yr for fixed opacity. Alternatively, for fixed accretion rate 10$^{-8}$ M$_{\odot}$/yr, r$_{\rm ice}$ varies from 1.2 AU to 4.8 AU across the range of $\kappa$ employed in different works (Min et al.\ 2011 and Davis 2005a). Yet, while $r_{\rm ice}$ can vary considerably based on different inputs, it is relatively 
straightforward to calculate once those inputs are fixed. 

In addition to fixing the location of the snow line, models try to determine
the concentration of H$_2$O vapor (the molar ratio of H$_2$O gas to H$_2$), 
inside and just outside $r_{\rm ice}$, as well as the concentration of H$_2$O ice particles
outside and just inside $r_{\rm ice}$. Higher ice/gas ratio over average beyond the snow line can encourage planetesimal growth. For example, the streaming instability is sensitive to solids-to-gas ratio (Youdin \& Goodman 2005). Higher vapor/gas ratio within the snow line may provide a more oxidizing environment for chemistry. Whether vapor is enhanced interior to the snow line or ice is more enhanced exterior to the snow line depends on the relative rates of evaporation of ice and diffusion of vapor, and freezing of vapor onto icy solids and rate of drift of these solids beyond the snow line.  Stevenson \& Lunine (1988) first argued that solids outside of the snow line would act as cold trap for water vapor diffusing from the inner nebula outward through the snow line. They argued that this would in time dehydrate the inner nebula, but also provide more solid material to enhance the growth of planetesimals beyond the snow line by a factor of $\sim$ 75, aiding formation of planets (and likely Jupiter) at the snow line, presumably at 5 AU. Cuzzi \& Zahnle (2004) later performed semi-analytical calculations for volatile transport and argued
that different inputs could give rise to a variety of outcomes; besides the possibility outlined
by Stevenson \& Lunine (1988), there could be a time in the disk's evolution when the inner disk
was enhanced in water vapor, if the flux of inward-drifting icy particles were higher than the 
outward flux of diffusing water vapor.
The inner disk could then become depleted in water vapor at later times, as the inward flux of 
drifting particles decreased (among other things, they are accreted into larger planetesimals). This was later verified by the numerical simulations of Ciesla \& Cuzzi (2006) who included the transport of vapor, dust, fast migrators and immobile asteroids within a disk evolution model and tracked their transport with time. Some models also consider that icy solids may be made of up multiple silicate particles held together by ice so that evaporation of ice can lead to the release of multiple small grains, dynamically coupled to the gas (Ida \& Guillot 2016; Schoonenberg \& Ormel 2017). While, the enhancements and/or depletions of vapor/ice inside and outside of the snow line affect the planet forming potential and the chemistry near the snow line, their effects also reach far beyond it as the snow line does not remain static but moves significantly with time, as noted above with disk models assuming different mass accretion rates. 

The relative rates of transport processes that produce variations in radial water abundance is strongly dependent on the turbulent viscosity parameter $\alpha$, which regulates both the diffusion of vapor within the snow line and diffusion and drift of solids beyond the snow line.  This can be seen by computing the timescale for volatile diffusion, 
$t_{\rm diff} \approx r^2 / {\cal D}$ $\approx r^2 / \nu$, 
where $r$ is the distance from the star, ${\cal D}$ is the diffusion coefficient, 
which is assumed to be ${\cal D} \approx \nu$, volatile viscosity.
The drift velocity of solids, which depends on the disk density is also
is greatly affected by $\nu$. 
Both diffusion and drift depend on the sound speed $c_{\rm s}$ and therefore temperature, which
depends on viscous heating and therefore also the turbulent viscosity $\nu$.
Moreover, as we argue below, it is not just the magnitude of $\nu$ that matters but also its 
spatial variation in the disk. 

\subsection{Turbulent Viscosity in Protoplanetary Disks}
Determining the extent of turbulence in protoplanetary disks is key to understanding the various processes of disk evolution and planet formation, as well as determining their associated timescales. Knowledge of the turbulence in disks provides insight into mass accretion rates and disk dissipation timescales (Lynden-Bell \& Pringle 1974; Hartmann et al.\ 1998), the rates of solid and volatile transport processes such as vertical settling of dust particles (Dullemond \& Dominik 2004), turbulent concentration of these particles to grow into larger particles (Cuzzi et al.\ 2003), radial diffusion of vapor (Cuzzi \& Zahnle 2004; Ciesla \& Cuzzi 2006; Desch et al.\ 2017) radial diffusion and drift of particles (Estrada et al.\ 2016; Desch et al.\ 2017), as well as the rates of planet growth by pebble accretion  (Xu et al.\ 2017) and final pebble isolation mass (Ataiee et al.\ 2018). The $\alpha$ parameter enters the disk evolution equations as the widely-used Shakura-Sunyaev (1973) parameterization of disk viscosity $\nu$. Here, $\nu$ is assumed to be a fraction of the product of the maximum velocity ($c_s$, i.e., local sound speed) and size ($H$, i.e., disk scale height) scales of turbulent eddies that could mix material in the disk (i.e., $\nu = \alpha c_s H $). This prescription has been useful to formulate turbulence in disk evolution even without the knowledge of the physical mechanism behind it.  

One possible constraint on $\alpha$ emerges from primitive material in the solar system. 
The size distribution of chondrules present in various chondrites are similar to a log-normal distribution around a mean size suggesting that they may have been aerodynamically sorted. A leading idea that could yield such a sorted distribution is turbulence (Cuzzi et al.\ 2001). If this is the case, then $\alpha$ can be inferred to be $\sim 10^{-4}$ at distances of $\sim$ few AU from the sun at around 1-3 Myr, thus providing a region- and time-specific reference point for $\alpha$ for the solar nebula (Cuzzi et al.\ 2001; Desch et al.\ 2007).

Observations of other protoplanetary disks have placed some more constraints on $\alpha$.
Hartmann et al.\ (1998) considered the mass accretion rates onto T Tauri stars and 
the viscous spreading of their disks over times, concluding that $\alpha \sim 10^{-2}$ 
represented a typical spatially and time-averaged value of $\alpha$.
Other studies reach similar conclusions: Hueso \& Guillot (2005) found 
$\alpha = 10^{-3} - 10^{-1}$ for DM Tau and $4 \times 10^{-4}$ to $10^{-2}$  for GM Aur;
Andrews et al.\ (2009, 2010) found $\alpha \approx 5 \times 10^{-4}$ to $8 \times 10^{-2}$
for disks in Ophiuchus.
More recently, the Atacama Large Millimeter Array (ALMA) has been used to constrain the 
turbulent broadening of CO emission lines in the outer regions ($> 30$ AU) of resolved disks,
finding $\alpha < 10^{-4} - 10^{-3}$ (Hughes et al.\ 2011; Guilloteau et al.\ 2012;
Simon et al.\ 2015; Flaherty et al.\ 2015, 2017, 2018; Teague et al.\ 2016).
These authors consider these values to be an upper limit, and low values of $\alpha$
are corroborated by other studies. 
Pinte et al.\ (2016) and recently, Dullemond et al.\ (2018) estimated $\alpha \sim 10^{-5} - 10^{-4}$ from the lack of smearing of
concurrent dust rings and gaps in observed disks.
Observational surveys of disks in the Lupus star-forming region find disks too compact to
have been viscously spreading (Ansdell et al.\ 2018).
Rafikov (2017) argued that the lack of correlation between $\alpha$ (determined from mass
accretion rates onto the central star) and other disk properties (e.g., disk mass, size, 
surface density, stellar mass, radius, or luminosity) means that turbulent viscosity cannot
be the major driver of disk evolution, and therefore $\alpha$ must be low.
In combination, these observations suggest $\alpha \approx 10^{-4} - 10^{-3}$ in the 
outer regions of protoplanetary disks. However, constraining its exact value by observations has been difficult by itself. Attempting to detect any spatial variation in this parameter is more difficult, and requires insight from models.

Many disk models have considered magnetohydrodynamics (MHD) mechanisms as the cause of angular 
momentum transport, especially the magnetorotational instability (MRI) of Balbus \& Hawley (1991).
Under the assumption of ideal MHD, the MRI is expected to yield a uniform $\alpha$ with high value 
($\alpha \sim 10^{-2}$) throughout the disk (Balbus et al.\ 1998 Rev Mod Phys).
It has long been recognized, however, that protoplanetary disks are subject to non-ideal MHD
effects, especially Ohmic dissipation, which will suppress the MRI at the disk midplane over a 
large range of distances from the star (Jin 1996; Gammie 1996). 
The result would be 'dead zones' between a few times 0.1 AU and about 10 AU, in which 
$\alpha$ is very low (perhaps $10^{-4}$ or less)  
Additional effects of ambipolar diffusion (Kunz \& Balbus 2004; Desch 2004; Bai \& Stone
2011; Simon et al.\ 2013a,b; Gressel et al.\ 2015) and the Hall effect (Wardle \& Ng 1999;
Bai 2014, 2015; Lesur et al.\ 2014; Simon et al.\ 2015) have been considered as well.
Within 10 AU, ambipolar diffusion will suppress the MRI in the surface layers away from the
disk midplane.
The Hall effect depends on the orientation of the magnetic field relative to the
disk rotation, and can lead to more or less angular momentum transport (Bai 2014; Lesur et al.\ 2014;
Simon et al.\ 2015; see \S 2). 
In general, Hall effects suppress the MRI inside 10 AU except in regions of modest density in certain 
circumstances. 

The recognition from models that the MRI is efficiently suppressed in large regions of protoplanetary
disks, combined with observations of $\alpha$ in outer disks far lower than expected for MRI-active
disks, has led to a resurgence of disk evolution models including purely hydrodynamic instabilities,
or hydrodynamic instabilities in concert with magnetically driven winds. 
Hydrodynamic instabilities such as vertical shear instability, or VSI (Stoll \& Kley 2014;
Flock et al.\ 2017), convective overstability (Klahr \& Hubbard 2014; Lyra 2014), and 
Zombie Vortex Instability (Marcus et al.\ 2013, 2015) typically yield $\alpha \sim 10^{-4}$
throughout the disk (Malygin et al.\ 2017; Lyra \& Umurhan 2018).
The variation of $\alpha$ across the disk would depend on the local cooling timescale, and probably
would not be uniform across the disk. 
Magnetically driven winds are expected to act inside about 1 AU, augmenting $\alpha$ due to 
hydrodynamic turbulence (Suzuki et al.\ 2016).

Overall, theoretical models strongly suggest that $\alpha$ could vary across the disk.
In our modeling of snow lines we therefore adopt three different profiles of $\alpha(r)$.
We consider a uniform $\alpha$ case, but also a case due to the MRI acting in a disk with a dead zone,
and a case in which VSI is augmented by magnetic disk winds in the inner disk. These profiles of $\alpha(r)$ are graphically represented in Figure \ref{threediffprofiles}.




\subsection{Overview}
In this paper we investigate, for the first time, the effects of spatially varying $\alpha$ on the 
distribution of water and volatiles in protoplanetary disks.
The relative rates of diffusion and particle drift depend on the magnitude of $\alpha$, because 
the flux of water vapor transported by diffusion out beyond the snow line is proportional to $\alpha$.
But the flux of particles inward across the snow line depends on the density of gas, and tends to 
be higher for lower gas densities.
Lower gas densities are associated with high $\alpha$, which leads to mass rapidly accreting onto 
the star.
Both the inward drift of icy particles and the outward diffusion of water vapor increase with 
$\alpha$. 
The concentration of vapor inside the snow line and the concentration of icy solids beyond the 
snow line depend sensitively on the balance between these two rates.
Therefore it may matter whether $\alpha$ is decreasing or increasing with $r$ across the snow line.
The different possible physical mechanisms for angular momentum transport lead to different profiles
for $\alpha(r)$ which, we hypothesize, could affect the distribution of water across the snow line.

\begin{figure*}
\plotone{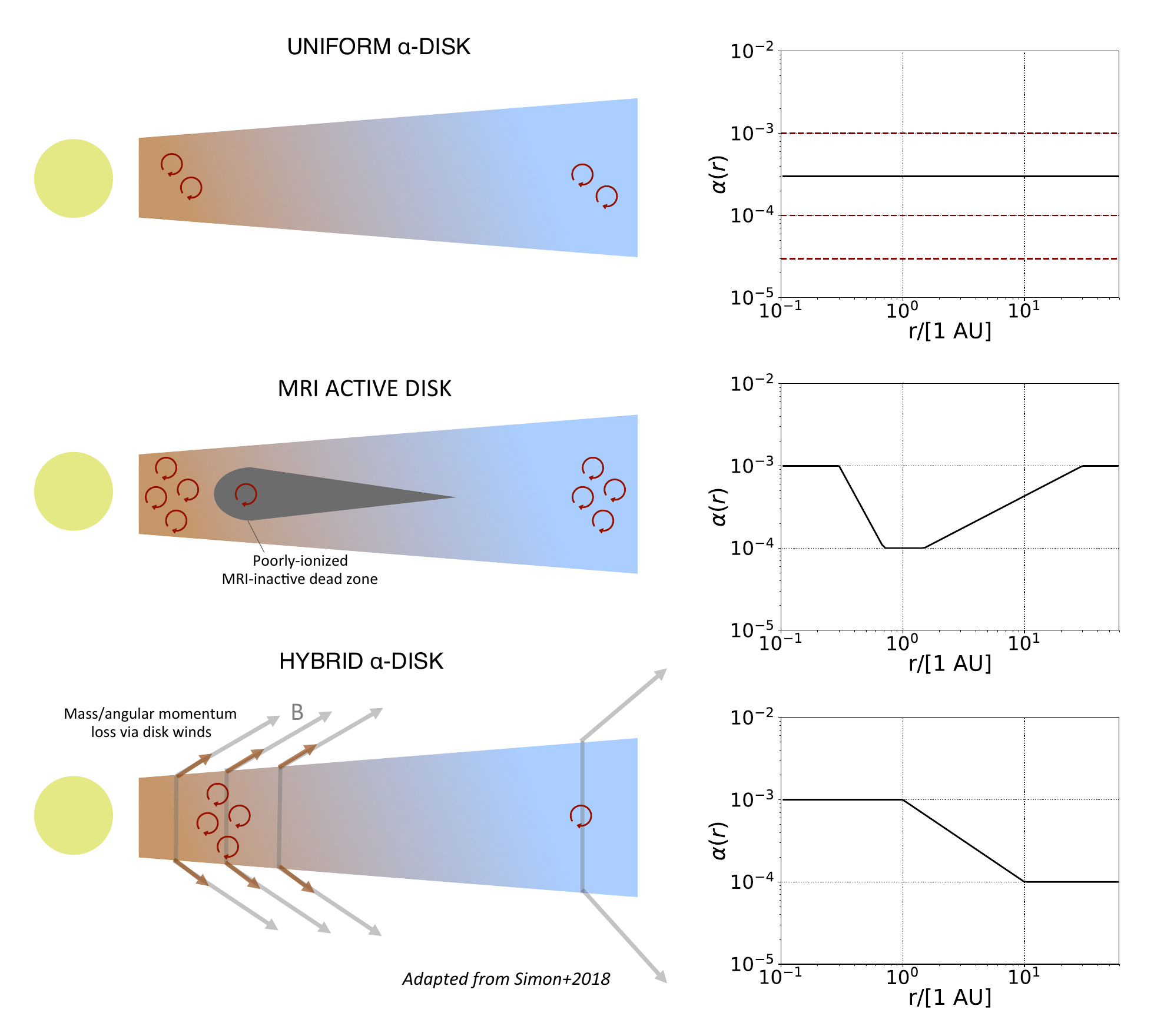}
\caption{The three $\alpha$ profiles considered in this work depicted with possible physical scenarios that would produce such a radial profile: a) the Uniform $\alpha$ profile, that is constant across radius; the range of constant $\alpha$ values explored in this study is also depicted as red-dashed lines;  ii) MRI-$\alpha$ profile, with an active innermost disk and a radially increasing $\alpha$ towards the outer disk; and iii) Hybrid $\alpha$ profile, a turbulent inner disk with radially decreasing $\alpha$ towards the outer disk. See \S2.2 for detailed discussion.}
\label{threediffprofiles}
\end{figure*}

We consider three distinct, physically motivated radial profiles of $\alpha$ (see Figure \ref{threediffprofiles}), and we
investigate the effect of each on disk evolution and the radial distribution of water across the 
snow line---and ultimately throughout the disk.
We examine how the distribution of water and volatiles depends not just on the overall value of $\alpha$ 
in a disk, but on the gradient of $\alpha(r)$ in a disk, especially across the snow line.
The paper is organized as follows. In \S 2 we describe the details of our code for calculating disk evolution
and volatile transport, and the assumptions underlying our snow line model. We also present our evolutionary models for the three different $\alpha$ profiles.
In \S 3 we describe the results of a suite of disk simulations assuming different $\alpha(r)$ and a parameter
study conducted to understand the effects of various assumptions about particle properties on the radial distribution of water in the protoplanetary disk.
In \S 4 we compare these numerical outcomes with theoretical and observational studies of disks, and with 
solar system data from meteorites.
We conclude that different profiles of $\alpha(r)$ yield subtle but distinct radial distributions of water
in protoplanetary disk.

\section{Methods}

\subsection{Disk Model}

\subsubsection{Structure and Flow of the Bulk Gas}

Our underlying disk model is described in detail in our previous paper (Kalyaan et al.\ 2015; 
hereafter Paper I). 
The 1(+1)D explicitly propagated finite-difference disk code features a protoplanetary disk of 60 AU in size, 
discretized into a logarithmic grid from 0.1 - 60 AU, and evolved using the canonical equations of 
Lynden-Bell \& Pringle (1974) of an axissymmetric viscously evolving disk.
In what follows we assume the disk is vertically well-mixed.

We consider a disk passively heated by starlight and actively heated by viscous dissipation, as described below. 
For passive heating, the temperature in the disk will vary with distance $r$ from the star as 
$T(r) \propto r^{-q_0}$, with $q_0 = 3/7$ (Chiang \& Goldreich 1997). 
The luminosity of the young Sun is considered to vary with time as per Baraffe et al.\ (2002), leading to a drop 
in temperatures over the first few Myr while the disk is present. 
In a passively heated disk, the temperature profile is
\begin{equation}
T_{\rm pass}(r) = 171.4 \, \left( \frac{t}{1 \, {\rm Myr}} \right)^{-1/7} \, 
 \left( \frac{r}{ 1 \, {\rm AU}} \right)^{-3/7} \, {\rm K}.
\end{equation} 
We also calculate the heating from viscous accretion using the results of the detailed 3D radiative transfer simulations 
of Min et al.\ (2011):
\begin{equation}
T_{\rm visc}(r) = \left[ \frac{27}{128} \, \frac{k}{\mu \sigma} \, 
 \Sigma(r)^2 \, \kappa \, \alpha(r) \, \Omega(r) \right]^{1/3},
\end{equation}
where $k$ and $\sigma$ are Boltzmann's constant and the Stefan-Boltzmann constant, $\mu = 2.33$ proton masses is 
the mean molecular weight, $\Sigma(r)$ is the surface density of gas, and $\Omega(r)$ is the Keplerian orbital frequency. 
We assume that a population of fine dust ($\sim 0.01-1 \mu m$) uniformly distributed and well-coupled to the bulk gas contributes to a radially-uniform and temperature-independent opacity of the disk material equivalent to $\kappa = 5 \, {\rm cm}^{2} \, {\rm g}^{-1}$ (as calculated by Semenov et al. (2003) for iron-rich composite aggregate grains) at 200-400 K. We note that the dust is a very small fraction ($< 1 \%$) of the overall mass of the gas.
Here the turbulence parameter $\alpha(r)$ is assumed to vary spatially according to the three cases described below in detail.
Note that it affects the temperature of the disk, in addition to affecting the gas and particle dynamics. 
To combine the effects of passive and active disk heating, we combine the two temperatures to get the total 
temperature, as follows:
\begin{equation}
T(r) = \left[ T_{\rm visc}(r)^4 + T_{\rm pass}(r)^4 \right]^{1/4}.
\end{equation} 
We neglect temperature-dependent variations in $\kappa$ at temperatures below silicate vaporization at
$\approx 1400$ K. Above that temperature the lack of opacity precludes a temperature gradient, so we assume a maximum
midplane temperature of $1400$ K.

\subsubsection{Transport of Vapor}

We consider water vapor to be a trace species in the bulk disk gas. 
We take the equations governing their evolution via advection and diffusion as follows 
(similar to those adopted by Clarke \& Pringle 1988, Gail 2001, Bockelee-Morvan et al.\ 2002; see Desch et al.\ 2017):
\begin{equation}
\frac{\partial \Sigma_{\rm vap}}{\partial t} = \frac{1}{2\pi r} \, \frac{\partial \dot{M}_{\rm vap}}{\partial r},
\end{equation}
where the mass flux of vapor is
\begin{equation}
\dot{M}_{\rm vap} = 2\pi r \, \Sigma \, {\cal D}_{\rm vap} \, \frac{\partial c}{\partial r}.
\end{equation}
Here $\Sigma_{\rm vap}$ and $\Sigma$ are the surface densities of vapor and bulk gas, 
$c = \Sigma_{\rm vap} / \Sigma$ is the concentration of the vapor, and 
${\cal D}_{\rm vap}$ is the diffusivity of the vapor. 
This does not necessarily equal the turbulent viscosity of the gas, $\nu$, but we take the ratio to be 
${\rm Sc} = \nu / {\cal D}_{\rm vap} = 1$, where ${\rm Sc}$ is the Schmidt number. 

In this work, we don't consider the effect of variation in Sc on $\alpha$ for gaseous species (we do however consider different Sc for solids as mentioned in later sections). Gaseous diffusivity or Sc will be dependent on the mechanism of turbulence as well as on the presence of a strong magnetic field, as studied by Johansen et al.\ 2006 (see Table 1 of their work). We also note that in order to simplify our computations, we also don't consider the vertical diffusivity of the tracer simply assuming that material is well-mixed with height. However, it need not be so and a 2-D model then will be necessary to understand the effects of varying vertical diffusion.

We use the donor cell method to convert mass flows from one grid zone to another.
At the inner boundary, we calculate the mass flux using the zero-torque boundary condition, as 
described in Paper I; 
if it is outward, we multiply by the concentration $c$ in the innermost zone to get the mass flux of vapor into 
the innermost zone.


\subsubsection{Transport of Solids}
Into this gaseous disk with vapor, we add small particles of 0.06 cm diameter, analogs to chondrules, which
are round, millimeter-sized particles found in abundance in chondritic meteorites (Scott \& Krot 2014; 
see their Table 1 for chondrule sizes).
In addition to diffusion and advection, intermediate-sized solid particles also drift relative to the gas,
usually inward, towards the star. 
This happens as a result of gas orbiting the star more slowly than the Keplerian orbital velocity $r \Omega$, 
by an amount $\eta \, r \Omega$, due to partial support by the pressure gradient force (usually outward).
Typically $\eta \sim 10^{-3}$. 
Particles orbiting at the Keplerian velocity feel a headwind that robs them of angular momentum, causing
them to spiral inward (e.g., Weidenschilling 1977). 
The rate at which particles drift inward depends on their Stokes number:
\begin{equation}
{\rm St} = \frac{\pi}{2} \, \frac{\rho_{\rm p} a_{\rm p}}{\Sigma},
\end{equation}
where $\rho_{\rm p}$ = 3 g cm$^{\rm -3}$ is the internal density and $a_{\rm p}$ is the radius of the particle.
This can be recast in terms of the aerodynamic stopping time of particles, $t_{\rm stop}$:
\begin{equation}
{\rm St} = \frac{1}{\Omega t_{\rm stop}}.
\end{equation}
In terms of the Stokes number, the drift velocity of particles with respect to the gas is 
\begin{equation}
V_{\rm drift} = \frac{ -{\rm St}^2 \, V_{\rm g,r} - {\rm St} \, \eta \, r \Omega }{ 1 + {\rm St}^2 },
\end{equation}
where $V_{\rm g,r}$ is the radial component of the gas velocity (Takeuchi \& Lin 2002).
This expression is valid in the Epstein regime, when $a_{\rm p} < \lambda$, where $\lambda$ is the
mean free path of gas molecules; this is typically a good assumption throughout much of a protoplanetary disk.
At 1 AU, particles of radius $\sim 30-100$ cm and ${\rm St} = 1$ can drift inward on timescales as short as 50 years,
but larger and smaller particles drift more slowly; 
chondrule-sized particles with radii 0.03 cm would typically take $10^5 - 10^6$ yr to drift inward 
(Weidenschilling 1977). We don't track the population of fine dust that produces the opacity of the disk material, and assume that it is well-coupled and homogeneously mixed throughout the disk.

The diffusion rate of particles also depends on their Stokes numbers.
Their diffusivity is 
\begin{equation}
{\cal D}_{\rm p} = \frac{ {\cal D}_{\rm gas} }{ 1 + {\rm St}^2 },
\end{equation}
where we assume the diffusivity of the bulk gas, like the vapor, is equal to the turbulent viscosity, $\nu$.

We apply the same boundary conditions on the flow of solids as we do on the flow of vapor. 

\subsubsection{Vaporarization \& Condensation}
Because our motivation is to calculate the spatial variation in water-to-rock ratio, we track the following
fluids independently: bulk disk gas; water vapor; ice-free or `rocky' chondrules (made of silicates);
`icy' chondrules that carry the mass of ice on particles the size of chondrules; ice-free, `rocky' asteroids 
(large silicate bodies too large to drift); and `icy' asteroids that carry the mass of ice on large bodies.
In reality, ice would coat the surfaces of rocky chondrules, slightly increasing their radius; in practice we
assume two populations of identical-size objects (chondrules or asteroids), one pure rock and the other pure ice.
We initialize the disk with a uniform concentration of vapor $c = 10^{-4}$, a uniform abundance of rocky chondrules
(with surface density $= 0.005 \times \Sigma$) and icy chondrules (also with surface density $= 0.005 \times \Sigma$),
and no icy or rocky asteroids. 

Ice in icy chondrules can convert to vapor if in a region warm enough. The rate at which this occurs depends on 
the local saturation water vapor pressure over ice, which is 
\begin{equation}
P_{\rm eq}(T) = 0.1 \, \exp \left( 28.868 - 6132.9 / T \right) \, {\rm dyn} \, {\rm cm}^{-2},  T > 169 \, {\rm K}
\end{equation}
from Marti \& Mauersberger (1993), and 
\begin{equation}
P_{\rm eq}(T) = 0.1 \, \exp \left( 34.262 - 7044.0 / T \right) \, {\rm dyn} \, {\rm cm}^{-2}, T \leq 169 \, {\rm K}
\end{equation}
from Mauersberger \& Krankowsky (2003).
The equilibrium vapor pressure relates to the surface density of water vapor as
\begin{equation}
\Sigma_{\rm vap,eq} = (2\pi)^{1/2} \, \left( \frac{ P_{\rm eq} }{ c_{\rm H2O}^2 } \right) \, H,
\end{equation}
where $H = C / \Omega$ is the scale height of the disk, $C$ the sound speed in the bulk gas, and $c_{\rm H2O}$ 
the sound speed in water vapor.
If $\Sigma_{\rm vap,eq}$ exceeds the total amount of water in icy chondrules and gas at radius $r$, we assume that
all of the water there is in vapor form; otherwise we assume $\Sigma_{\rm vap} = \Sigma_{\rm vap,eq}$ and assume
the remaining water is in the form of water ice (in icy chondrules). 

\subsubsection{Particle Growth}
To simulate the growth of particles into planetesimals, we assume a fraction of the chondrule population at each
radius $r$ is converted into asteroid bodies every timestep of the code.
Specifically, asteroids are presumed to grow on a timescale $t_{\rm grow}$, so that in a timestep $dt$ the mass
of chondrules is reduced by an amount $\Sigma_{\rm chon} (dt / t_{\rm grow})$, and the same mass is added to the
mass of asteroids. 
We assume typically $t_{\rm grow} = 1$ Myr. 
We do not include detailed models of fragmentation or growth of these particles into larger bodies, rather assuming
that a fraction of the mass grows per time interval into large bodies by growth mechanisms such as streaming instability
or pebble accretion (Johansen et al.\ 2008; Lambrechts \& Johansen 2014).
We do not model the size of asteroids, since bodies of any size more than a few km would take $\gg 1$ Myr to radially drift
(Weidenschilling 1977).
We ignore migration of asteroids by other mechanisms (e.g., dynamical resonances or scatterings). 

\subsection{Turbulence Radial Profiles}

We consider three different radial profiles for the turbulence parameter $\alpha(r)$ and examine the response
of water vapor and ice to each profile. 
A major goal of this paper is to study the effect of these different profiles---not just the magnitude of $\alpha$,
but the variation of $\alpha$ with distance $r$ from the star--- on the distribution of water in the disk.
The most consistent way to do this would be to adopt power law profiles $\alpha \propto r^{-a}$, with $a$ carrying
different slopes, but the value of $\alpha$ at the snow line remaining fixed.
The problem with this approach is that the different values of $\alpha$ inside and outside the snow line also
lead to different distributions of mass and $\Sigma(r)$ profiles. Because the temperature depends on accretional
heating and therefore $\Sigma(r)$, the location of the snow line would also vary. 
A different approach would be to adopt the profiles of $\alpha(r)$ predicted from first principles by different
theories.
The problem with this approach is that first-principles approaches to deriving $\alpha(r)$ are not especially robust or
predictive. They also would probably depend sensitively on inputs such as temperature, density, ionization levels, etc.
Improvements in the models, or just differing input assumptions, are likely to lead to different $\alpha(r)$ profiles. 
As a compromise, we consider three $\alpha(r)$ profiles that capture the flavor of different physical mechanisms.
Case I considers a uniform value of $\alpha$ throughout the disk.
Case II considers a disk subject to the MRI, with a dead zone and low $\alpha$ at intermediate radius, bracketed by 
larger $\alpha$ at other radii. 
Case III considers a disk subject to purely hydrodynamic instabilities yielding uniform $\alpha$, augmented by 
enhances transport due to magnetic disk winds in the inner disk.
Across the snow line region (typically at several AU), $\alpha$ is uniform in Case I, increasing with $r$ in Case II, 
and decreasing with $r$ in Case III. 
Below we present the three $\alpha(r)$ profiles used, and discuss their physical motivations and justifications. We also discuss the effect each $\alpha(r)$ has on the disk surface density and temperature across $r$. 

\begin{figure*}
\plotone{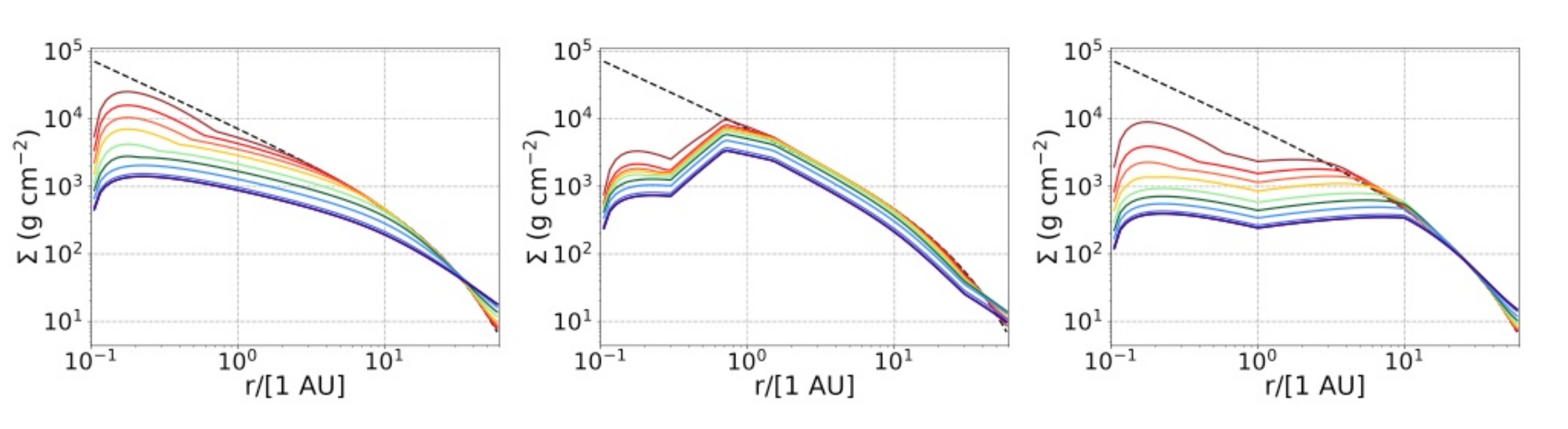}
\caption{Plots show the evolution of the surface density profiles $\Sigma(r)$ for the three disk evolution models considered in this work: a) the Uniform $\alpha$ profile, b) MRI-$\alpha$ profile, and c) Hybrid $\alpha$ profile. Note the structure of the disk driven by the MRI, with much of the mass concentrated at $\sim$ 1 AU region, as well as the structure of the hybrid $\alpha$ disk with largely constant $\Sigma$ up to 10 AU. Different colors represent different times: 0 (dashed), 20kyr (red), 50kyr (orange), 100kyr(yellow), 200kyr (light green), 500kyr (green), 1Myr (light blue), 2Myr (blue), 4Myr (dark blue), and 5Myr (violet).}
\label{sigmaprof}
\end{figure*}

\begin{figure*}
\plotone{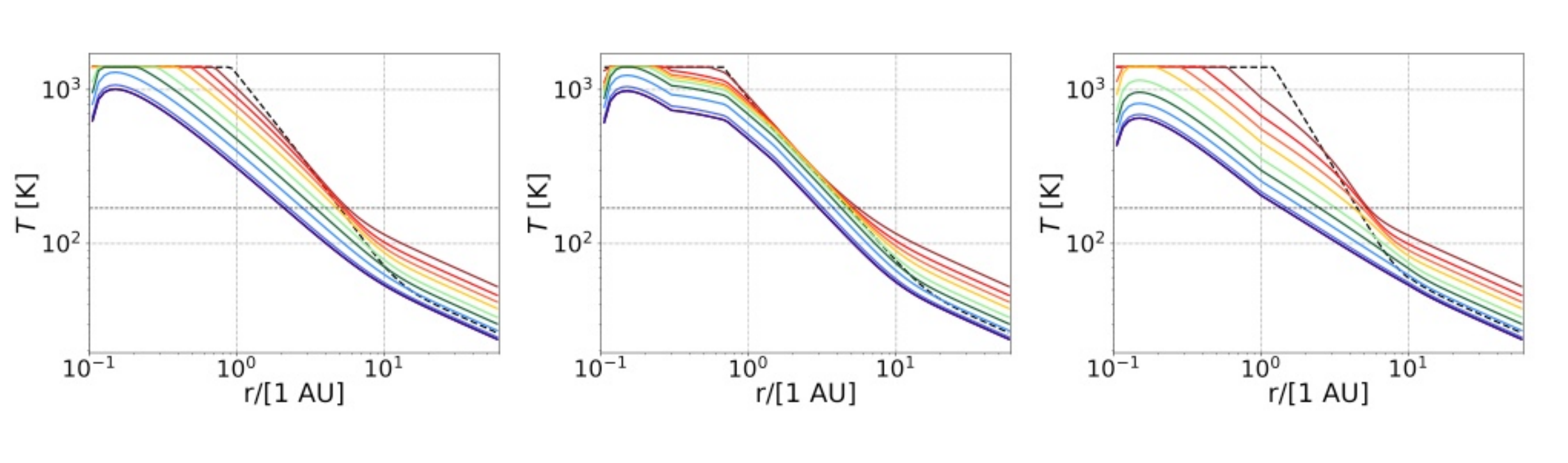}
\caption{Plots show the evolution of the radial temperature profiles $T(r)$ for the three cases: a) the Uniform $\alpha$ profile, b) MRI-$\alpha$ profile, and c) Hybrid $\alpha$ profile. Different colors represent different times: 0 (dashed), 20kyr (red), 50kyr (orange), 100kyr(yellow), 200kyr (light green), 500kyr (green), 1Myr (light blue), 2Myr (blue), 4Myr (dark blue), and 5Myr (violet).}
\label{tempprof}
\end{figure*}

\subsubsection{Uniform $\alpha$ profile}

The uniform $\alpha$ profile is one that is most commonly used in disk models as this profile makes the least assumptions regarding the specific physical mechanism that contributes to the viscosity in the disk material and transports angular momentum. As mentioned before, $\alpha$ features in the parameterization of viscosity $\nu$ as $\nu = \alpha c_s H$, where $c_s$ is the sound speed, and $H$ is the disk scale height; its choice decides how large turbulent eddies can be and how fast they can flow, and therefore, how efficiently material is mixed and transported around in the disk. Since the dominant physical mechanism behind momentum transport has been hard to determine and still requires more observational verification (see \S1), this profile remains attractive for wide use in disk models. We therefore define ``CASE I" as follows:
\begin{equation}
\alpha_{\rm I} = 3 \times 10^{-4} \, \, \, \, ({\rm all} \, \, r).
\end{equation}
We also consider a range of globally uniform values $\alpha$ between $10^{-5}$ and $10^{-3}$. 
This choice of $\alpha$ is motivated by the observations of protoplanetary disk evolution, but 
also by numerical simulations of hydrodynamic turbulence such as the VSI, 
and by the concentration of chondrules by turbulence (see \S 1.3). 
While continued simulations of VSI and other hydrodynamic instabilities may ultimately predict variations
in $\alpha$ with position, at the current time a uniform $\alpha$ throughout the disk is not inconsistent
with purely hydrodynamic turbulence. 

\subsubsection{MRI $\alpha$ Profile}
While the uniform $\alpha$ assumption is widely used and simple to implement, it is unlikely that a 
protoplanetary disk would have uniform $\alpha$. 
This is certainly true if angular momentum transport is dominated by the MRI, as it only operates
at full efficiency in regions that are sufficiently ionized.
The cold, poorly-ionized midplane regions from inside 1 AU, out to beyond 10 AU (``dead zones") are unlikely 
to be very MRI-active; however,
the regions close to the Sun may be MRI-active due to thermal ionization (Desch \& Turner 2015), and 
the regions far from the Sun may be MRI-active due to cosmic-ray ionization.
These effects alone introduce radial variations, as the innermost and outermost portions might have 
relatively high turbulence, which we arbtirarily take to be $\alpha \sim 10^{-3}$, but the dead zone
regions would have lower $\alpha$. 
In the dead zone regions, the low-density upper layers are likely to be ionized either to a uniform surface
density $\Sigma_{\rm a} \sim 100 \, {\rm g} \, {\rm cm}^{-2}$ (Gammie 1996) or to radially-dependent surface 
densities $\Sigma_{\rm a} < 10 \, {\rm g} \, {\rm cm}^{-2}$ due to X-ray ionization 
(Lesniak \& Desch 2012; Kalyaan et al.\ 2015).
The vertically averaged value of $\alpha$ would be $\alpha_{\rm a} \Sigma_{\rm a} / \Sigma$, where 
$\alpha_{\rm a}$ is the value in the active layer. 
In general the variations in $\Sigma$ and $\Sigma_{\rm a}$ with $r$ are likely to lead to radial variations of 
$\alpha_{\rm a}$ within the active layer as well.
In order to understand the structure of a disk evolving with the MRI, Kalyaan et al. (2015) employed formulations from Bai \& Stone (2011) for MRI accretion which considered the non-ideal MHD effects of ambipolar diffusion. The model disk was assumed to be ionized by radially dependent X-rays from the protostar, ambient cosmic rays, as well as an internal ionizing source of short-lived radionuclides. A simple chemical model was assumed that included both recombination of ions and electrons in the gas phase as well as grain surfaces. $\alpha$ was determined from the local ion density; a vertically-averaged $\alpha(r)$ profile therefore assumed the shape of a power law with a positive slope, i.e., $\alpha$ increased with $r$, as the disk became more and more MRI-active (due to increasing ion density) with increasing $r$. The only exception was the innermost disk ($<$ 0.3 - 1 AU), which is likely to have temperatures exceeding $\sim$ 1000 K. In the limit of ideal magnetohydrodynamics, high temperature ionization processes such as the collisional ionization of alkali metals (Armitage 2011) as well as thermionic emission of dust grains (Desch \& Turner 2015) render the entire vertical extent of the innermost region MRI-active. Therefore, neglecting some non-ideal effects, the shape of the MRI-$\alpha$ profile as shown in Figure \ref{threediffprofiles} seems reasonable to assume. The vertically averaged $\alpha(r)$ value calculated in Kalyaan et al.\ (2015) varied strongly with $r$ as 1 $\times 10^{-5}$ in the inner disk within 1 AU to 0.01 in the outer disk beyond 20 AU, rather than the one-order of magnitude difference between the lowest and highest $\alpha$ assumed here. 

The second $\alpha$ profile we consider (``CASE II"), motivated by studies of the magnetorotational instability (MRI), 
therefore is given by 
\begin{equation}
\alpha_{\rm II}(r) = \left\{
\begin{array}{llc} 
1 \times 10^{-3},                                                   & r \leq 0.3 \, {\rm AU} \\ 
1 \times 10^{-4} \, \left(  r / 0.3 \, {\rm AU} \right)^{-2.718},   & 0.3 \, {\rm AU} < r \leq 0.7 \, {\rm AU} \\ 
1 \times 10^{-4},                                                   & 0.7 \, {\rm AU} < r \leq 1.5 \, {\rm AU} \\ 
1 \times 10^{-4} \, \left(  r / 1.5 \, {\rm AU} \right)^{+0.769},   & 1.5 \, {\rm AU} < r \leq 30 \, {\rm AU} \\ 
1 \times 10^{-3},                                                   & 30 \, {\rm AU} < r 
\end{array} 
\right.
\end{equation} 
We note that this profile is consistent with $\alpha \approx 3 \times 10^{-4}$ in an averaged sense.
It equals $3 \times 10^{-4}$ at 6.2 AU. At 3 AU, $\alpha = 1.7 \times 10^{-4}$. 

To verify the consistency of this profile with results of recent studies, we note that Flaherty et al.\ (2015,2017) find an upper limit of turbulent $\alpha \sim 10^{-3}$ in the outer disk ($>$ 30 AU) of HD163296. Bai (2015) present simulations that yield $\alpha$ $\sim$ 10$^{-4}$ (10$^{-6}$) if $\Omega B > 0$ ($\Omega B < 0$) at 5 AU. We assume a scenario where turbulent transport begins to gradually increase in efficiency towards larger radii ( $\sim$1 AU $< r <$ $\sim$30 AU). Beyond 30 AU, we assume turbulent $\alpha$ reaches 1 $\times$ 10$^{-3}$ an order of magnitude higher than the inner disk. 
Despite these uncertainties, we feel our case II profile captures a key attribute of the MRI, which is that it
should be less active in dead zone regions, with $\alpha$ increasing in magnitude with increasing $r$ beyond the
dead zone. 

\subsubsection{Hybrid $\alpha$ profile}

At the current time, models invoking hydrodynamic instabilities such as VSI, would predict a low level of 
turbulence acting throughout this disk, with $\alpha \sim 10^{-4}$; these models are not developed to the point
that radial variations can be strongly argued for. 
A disk evolving purely by non-magnetic, hydrodynamic instabilities such as the VSI is probably represented well
by a uniform $\alpha$ disk as in our Case I profile.
Disks evolving by magnetic instabilities such as the MRI are probably characterized by our Case II profile.
The third example we consider is a disk that is evolving primarily by hydrodynamic instabilities such as VSI,
but augmented by magnetically controlled angular momentum transport, not through the MRI but through disk winds.

For example, Simon et al.\ (2018) explain the low levels of turbulence measured by Flaherty et al.\ (2015, 2017)
in the outer portions of some disks, by invoking attenuation of cosmic rays by strong winds launched from the inner disk.
These winds, in turn, would be generated by magnetocentrifugal outflows relying on large-scale magnetic fields.
Suzuki et al.\ (2016) have recently performed global numerial simulations of disk winds, considering the global
energy budget of a protoplanetary disk. 
Suzuki et al.\ (2016) find that the disk winds launched by the outflows would induce a torque in the disk
characterized by $\alpha_{\phi,z} \sim 10^{-4}$. This is not the same as the traditional $\alpha$ parameter, 
which relates directly to the $\alpha_{r,\phi}$ component of the Reynolds or Maxwell stresses, but it plays
a similar role in that the disk wind would drive angular momentum transport, lead to heating (by ambipolar 
diffusion) and the Reynolds stress could help particles diffuse. 
Other simulations find a role for the MRI in the upper layers of the disk in driving the disk winds.
Simulations by Bai et al.\ (2015) find $\alpha \sim 10^{-6} - 10^{-4}$ at 5 - 15 AU, depending on whether
the magnetic field is aligned or anti-aligned with the disk's rotation. 

These results suggest a hybrid model in which the disk overall is characterized by low levels of 
$\alpha \sim 10^{-5}$ throughout, except in the inner disk, where $\alpha > 10^{-4}$ may obtain due to magnetic
disk winds, possibly in concert with the MRI. 
Such a profile was recently considered by Desch et al.\ (2018) in their disk evolution model to explain the
abundances of CAIs (calcium-rich, aluminum-rich inclusions) and refractory elements in different meteorite types.
This model demands a low level of $\alpha \sim 10^{-5}$ in the outer disk ($> 10$ AU) to prevent mixing of gas
with the inner disk, which is depleted in refractory elements; higher values of $\alpha$ would not allow CI 
chondrites to chemically match the solar photosphere. 
The model also demands a higher value of $\alpha$ in the inner disk ($< 1$ AU), up to $5 \times 10^{-4}$, so
that CAIs will be efficiently transported outward in the disk; lower values of $\alpha$ would not lead to 
efficient transport of CAIs to the carbonaceous chondrite-forming region, but higher values of $\alpha$ would 
drain the inner disk of material before ordinary and enstatite chondrites could form. 

Based in part on the success of the Desch et al.\ (2018) model, and in part on the physical plausibility of 
VSI augmented by disk winds, the third $\alpha$ profile we consider (``CASE III") is given by 
\begin{equation}
\alpha_{\rm III}(r) = \left\{
\begin{array}{llc} 
1 \times 10^{-3},                                                   & r \leq 1  \, {\rm AU} \\ 
1 \times 10^{-3} \, \left(  r / 1 \, {\rm AU} \right)^{-1},         & 1 \, {\rm AU} < r \leq 10 \, {\rm AU} \\ 
1 \times 10^{-4},                                                   & 10 \, {\rm AU} < r 
\end{array} 
\right.
\end{equation} 
We note that like the case II profile, the typical value of $\alpha$ in the case III profile is $\approx 3 \times 10^{-4}$, 
in an average sense, and equals that value at 3 AU. 

\subsubsection{Effect of Different $\alpha(r)$ on Disk Structure and Evolution}

A radially-varying $\alpha$ profile considers a disk with a variable radial efficiency of mass transport. Such a disk will see more rapid mass transport of the bulk gas  or larger variations in local density in its more turbulent regions (i.e., higher $\alpha$) over time than its less turbulent and therefore quiescent regions. This yields distinctly different structure than the smooth $\Sigma$ profiles of the traditional uniform $\alpha$ disk (e.g. Hartmann et al.\ 1998). In this work, we consider two variable $\alpha$ evolutionary disk models: i) MRI-driven disk; and ii) Hybrid $\alpha$ disk.

The case of an MRI-driven disk (CASE II) was investigated in detail by Kalyaan et al.\ (2015), where we assumed an MRI-active inner disk, and an outer disk that was increasingly turbulent with heliocentric distance due to decreasing gas density and increased ionization from cosmic ray penetration. Here we adopt a time-averaged parameterized form of the same $\alpha$ profile (i.e., CASE II). This $\alpha(r)$ yields a tenuous inner disk and a massive central region that results in mass pile-up from the outer disk. The resulting structure from such an $\alpha$ profile closely resembles that of the structure of transition disks as pointed out by Pinilla et al.\ (2016).

We also consider another radially varying $\alpha$ profile (Case III) where we assume higher turbulent $\alpha$ in the inner disk that gradually decreases with increasing $r$. Such a profile yields a distinctly different structure where $\Sigma(r)$ is flat throughout the inner disk upto $\sim$ 20 AU. Beyond this, $\Sigma(r)$ exponentially drops, varying little from the initially assumed profile.

Two varying $\alpha$ profiles thus produce two different disk structure, and yield different evolutionary timescales. It is in this context that we consider the transport of tracer volatiles and solids that are mixed within the bulk gas to varying extents according to the local $\alpha(r)$ and study its effects.

\section{Results}
In this work, we carry out a suite of simulations of an evolving disk with volatile transport under different prescriptions of turbulence. We explore a range of globally uniform $\alpha$ values as well as radially varying $\alpha$ profiles that are assumed to be motivated by different mechanisms of angular momentum transport. The results of these simulations are complex and interdependent on several parameters. Therefore, we also perform a set of simulations where we only change one parameter at a time to tease out important trends while keeping the other parameters constant, as many of these effects occur concurrently in the radially varying $\alpha$ simulations. First, we discuss our canonical uniform $\alpha$ disk in detail. We then vary some parameters (namely, size of the drifting particle, diffusivity of vapor and small solid particles, opacity $\kappa$ of the disk gas and finally the timescale of growth of asteroids) that regulate properties of either particles, vapor or the general disk material/processes. Finally, we explore volatile transport in disks with different global $\alpha$ and radially varying $\alpha$ profiles. 

\subsection{Our Canonical Uniform $\alpha$ case}
\begin{figure*}
\plotone{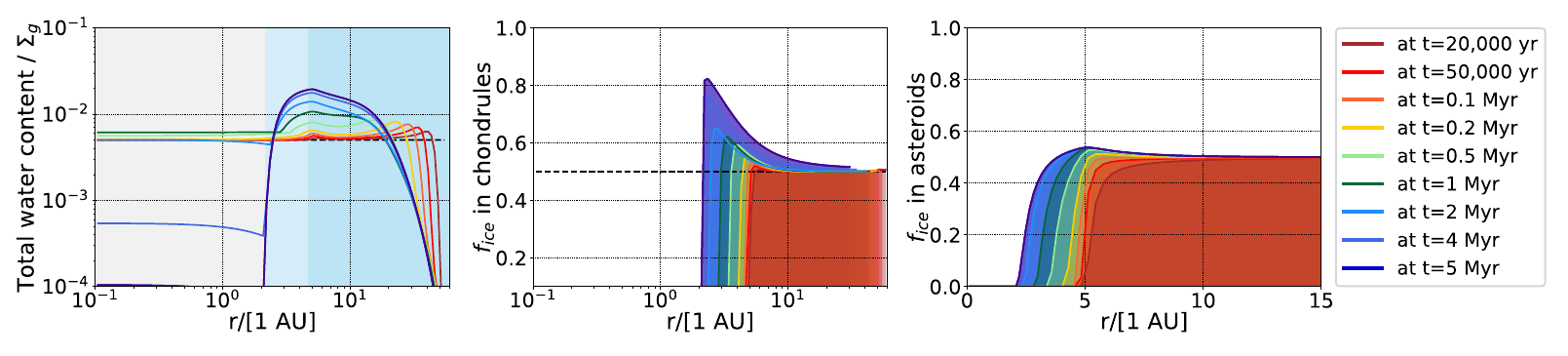}
\caption{Plots show the results of our canonical uniform $\alpha$ case, i.e, CASE I. Here, $\alpha$ is considered to be 3 $\times$ 10$^{-4}$ at all $r$. (a) shows total water content/$\Sigma_{g}$ with $r$, where total water = $\Sigma_{\rm vap}+\Sigma_{\rm chon}+\Sigma_{\rm ast}$; (b) shows radial variation in the water-rock ratio in chondrules; (c) shows radial variation in the water-rock ratio in asteroids that grow from chondrules.}
\label{unialpha}
\end{figure*}

We first discuss the results of our canonical case of the uniform $\alpha$ disk in detail, where $\alpha$=3 $\times$ 10$^{-4}$ at all $r$, which is shown in Figure \ref{unialpha}. The figure comprises of three plots that help us trace the water content in the disk in different useful ways. In the figure, the first plot (Figure \ref{unialpha}a) shows the radial distribution of total abundance of water in ice and vapor, i.e., sum of $\Sigma_{\rm vap}$ and $\Sigma$ of ice in solids (chondrules + asteroids)/$\Sigma_{\rm gas}$ at each $r$ at different times ranging from t = 20,000 yr - 5 Myr. The location of the snow line is evident as the radius at 2 AU at which this quantity changes abruptly at each timestep. The dark blue region denotes radii always beyond the snow line and is composed of only ice between times shown (20000 yr - 5 Myr). The gray shaded region denotes the disk radii always within the snow line that is composed of only vapor throughout the course of our simulation up to 5 Myr.  The light blue region between them denotes regions that start with H$_2$O as vapor, but later see H$_2$O only as solid ice as the snow line moves inward as accretional heating diminishes. Figure \ref{unialpha}b shows the fraction of the mass in small particles that is ice, at different radii $r$ and times $t$. 
Far out in the disk, the particles assume cosmic abundances and are assumed to be 50\% ice and 50\% rock. 
Inside the snow line, no ice exists in small particles.  
Just beyond the snow line, water can be cold-trapped on small particles, which see an enhancement in the ice fraction
above cosmic abundances. 
These small particles are subject to rapid radial drift and are also the precursor materials for asteroidal bodies that
grow directly from them, e.g., by streaming instability. 
Asteroidal material at a given location  is assumed to capture at any time in the disk a fraction of the rocky chondrule 
mass and the icy chondrule mass at each location. These asteroids are assumed not to radially migrate after that. 
Figure \ref{unialpha}c shows the fraction of asteroidal material at various radii $r$ and times that is ice.
Note that the radial scale is linear, not logarithmic as in Figures \ref{unialpha}a and \ref{unialpha}b.
At early times, asteroid material has cosmic abundances of water (50\% ice fraction) outside the snow line at about 5 AU,
and no ice inside that. As the snow line moves inward, more asteroidal material between 3 and 5 AU can acquire water ice. 

Figure \ref{unialpha} illustrates several processes shaping the distribution of volatiles across the snow line and throughout 
the disk. The concentration of vapor in the inner disk depletes with time as it is either diffused inward and accreted onto the star, or is back-diffused through the snow line to deposit as more ice on solids (Figure \ref{unialpha}). We also have to note that vapor diffusion is accompanied by diffusion of chondrules in two ways through the snow line: as inward diffusion of icy chondrules, and as both inward \textit{and} outward diffusion of silicate chondrules (holding to our assumption that ice on chondrule evaporates instantaneously upon approaching the snow line). Silicate chondrule diffusion inward from the snow line to the inner nebula however is counteracted by the outward silicate diffusion towards the outer nebula that have equal rates. Therefore, silicate diffusion through the snow line does not enrich or reduce the abundance of ice beyond the snow line. Effectively, it is only the vapor that is back-diffused through the snow line to the outer nebula, which piles up as more ice with time just beyond the snow line. This leads to increase in the fraction of water-ice in chondrules from initial concentration of 0.5 (assumed from equal abundances of ice and rock) to peak abundance of 0.8 just beyond 2.0 AU at 5 Myr (Figure \ref{unialpha}). The peak abundance of ice beyond the snow line eventually forms the radially smeared-out peak abundance in the ice-fraction of asteroids in Figure \ref{unialpha}c. The radial smearing occurs as the snow line moves inward and the chondrules and the ice distribution moves inward, leaving the asteroids that form in their wake to remain in place. 

\begin{figure*}
\epsscale{0.67}
\plotone{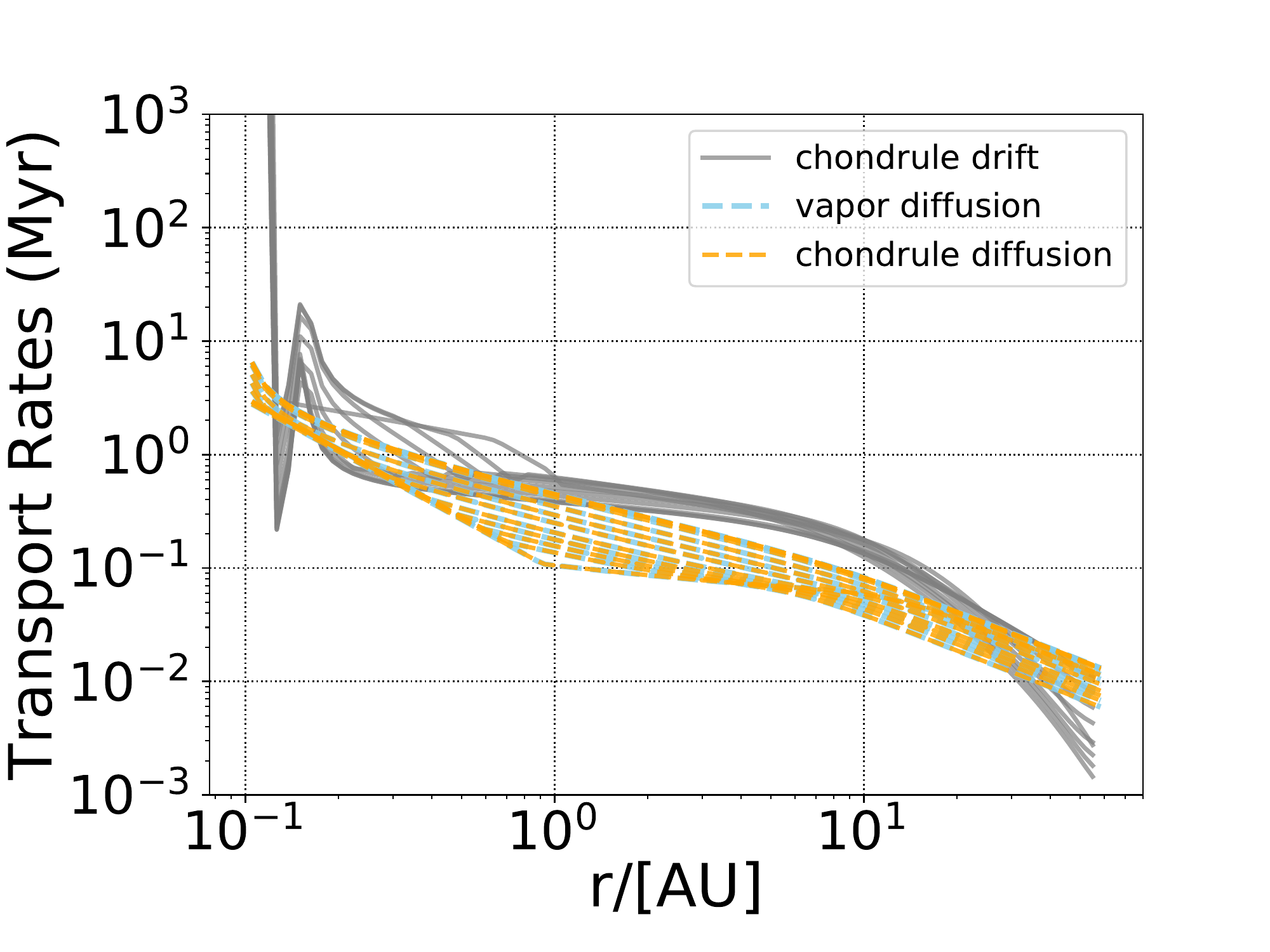}
\caption{Plot shows the timescales of the various radial transport processes of volatiles for the uniform $\alpha$ case as depicted in Figure \ref{unialpha}. The gold dotted lines show the radial variation in chondrule diffusion rates at the simulation times plotted in all other plots (20,000 yr - 5 Myr); the blue dotted lines (overlapped by the gold) show the radial variation in vapor diffusion rates; and grey lines show the radial variation in drift rates of chondrules. All transport rates computed for a 1 AU region. We note that on approaching the snow line, drift rates of icy chondrules becomes irrelevant as they are sublimated to vapor, and beyond the snow line, vapor is immediately converted into ice on chondrules, rendering diffusion rates of vapor here invalid. It is nevertheless shown here for completion and because these rates are still valid for silicate chondrules.}
\label{unialphatimescale}
\end{figure*}

To facilitate comparisons to this canonical case, in Figure \ref{unialphatimescale} we present the timescales of different transport
mechanisms in the disk, defined as $2\pi r \, \Sigma / (\partial \dot{M} / \partial r)$, where $\Sigma$ and $\dot{M}$ 
refer to the relevant fluid (chondrules, or H$_2$O, etc.).
Throughout much of the disk, the diffusion of chondrules and vapor are equally rapid and the most rapid processes in 
the disk, with typical timescales $\sim 10^5$ yr. Chondrule drift is almost as rapid, with associated timescales
$\sim 3 \times 10^5$ yr.

\subsection{Parameter Study of a Uniform $\alpha$ Disk}
We perform a parameter study for some of the most important input parameters in our disk model that is able to significantly influence the water distribution across the disk. We performed simulations for a range of chondrule sizes, Schmidt number, timescale for growth of asteroids and opacity of disk material. These parameter studies were performed with the canonical uniform $\alpha$ disk with global $\alpha$ = 3 $\times$ 10$^{-4}$ for comparison. The results of this parameter study are highlighted in Figures \ref{paramachon}, \ref{paramsc}, \ref{paramtgrowth}, \ref{paramkappa}, and \ref{alltimescales} and discussed in detail below. 

\subsubsection{Effect of varying chondrule size}

\begin{figure*}
\plotone{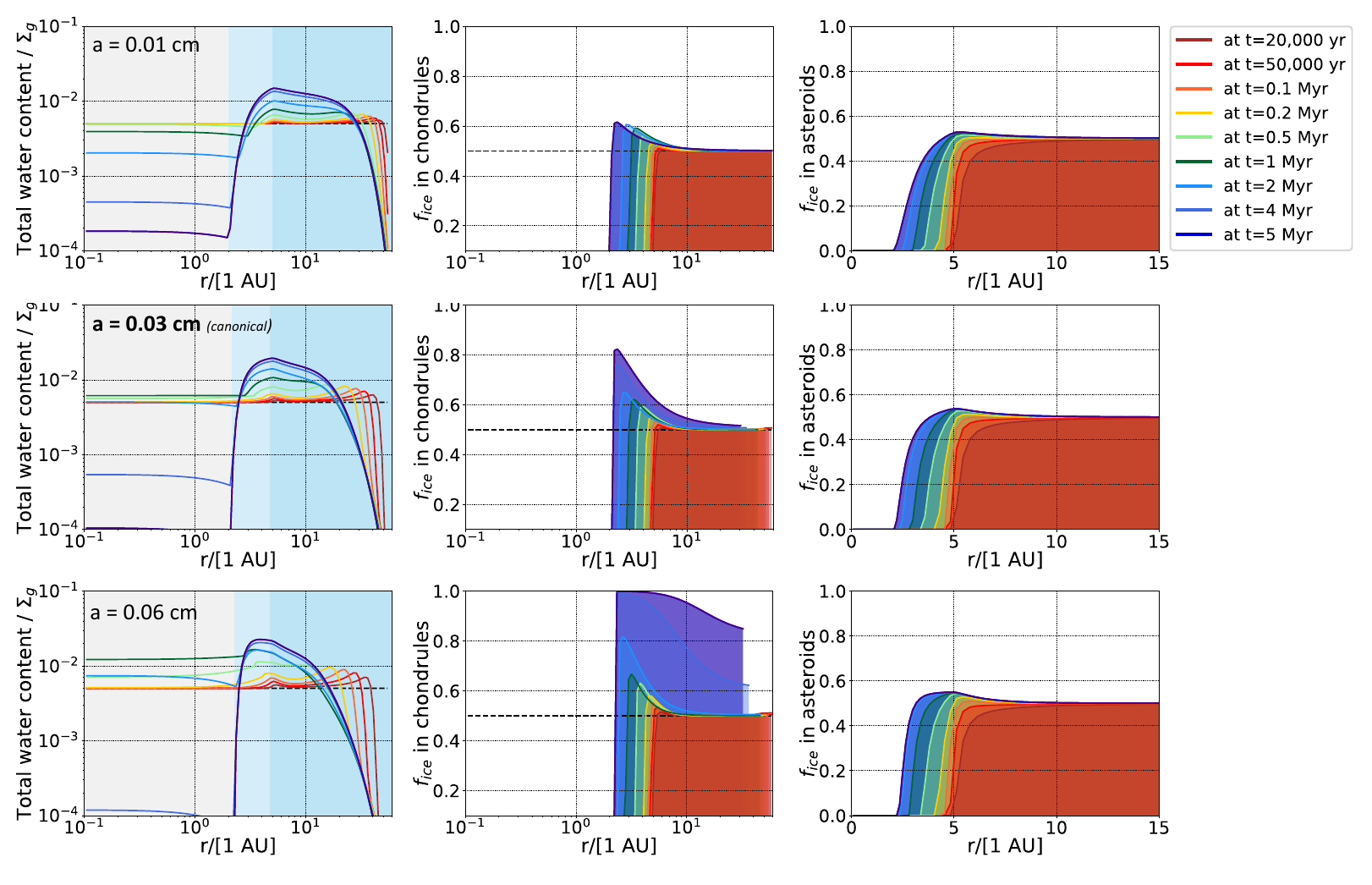}
\caption{Plots show results of variation of parameter chondrule size $a$ in an uniform $\alpha$ disk. Increasing chondrule size leads to higher chondrule ice-rock ratios beyond the snow line; an indirect effect of more rapidly drifting particles bringing in more water to the inner nebula which is then back-diffused through the snow line to refreeze as ice. Figures and colors same as previous plots.}
\label{paramachon}
\end{figure*}

In this study, the chondrule radius was varied from 0.01 cm, to 0.03 cm (our canonical case), to 0.06 cm, to explore 
the changes in water distribution due to different chondrule sizes.  
Increasing the chondrule radius has a dramatic effect on the water distribution, as seen in Figure \ref{paramachon}. 
With increasing chondrule size, the ice fraction of chondrules just beyond the snow line increases greatly; at 5 Myr
it is 60wt\% for $a = 0.01$ cm, 80wt\% for $a = 0.03$ cm, and nearly 100wt\% for $a = 0.06$ cm. 
We attribute this to a change in the relative rates of diffusion and drift. 
The diffusion rate of particles can vary with particle size through the Stokes number (Equation 9), but in the limit
${\rm St} \ll 1$, appropriate for chondrules, the diffusivity does not appreciably vary with particle size. 
The drift rate, on the other hand, is proportional to ${\rm St}$ and therefore particle size.
Larger chondrules diffuse as quickly as small chondrules, but large chondrules drift more rapidly; for the large
chondrules the drift rates approach the diffusion rates. 
Whatever water vapor diffuses outward across the snow line is more rapidly brought back across the snow line into the 
inner disk; but, simultaneously, drift from the outermost portions of the disk is more rapid, and more water overall
is brought to the snow line. 
This has the additional effect of limiting the depletion of water vapor from the inner disk inside the snow line.
While the inner disk content steadily is more depleted over time in the case with $a = 0.01$ cm, the case with $a = 0.03$ cm
has roughly constant water vapor in the inner disk for about 0.5 Myr, implying the inner disk is temporarily `flooded' with 
H$_2$O. 
For the case with $a = 0.06$ cm, the vapor abundance is enhanced for 1 Myr before rapidly depleting. 
Because $\alpha$ is the same in all three cases, the snow line is at the same location for all three cases. 

\subsubsection{Effect of varying particle diffusivity}

\begin{figure*}
\plotone{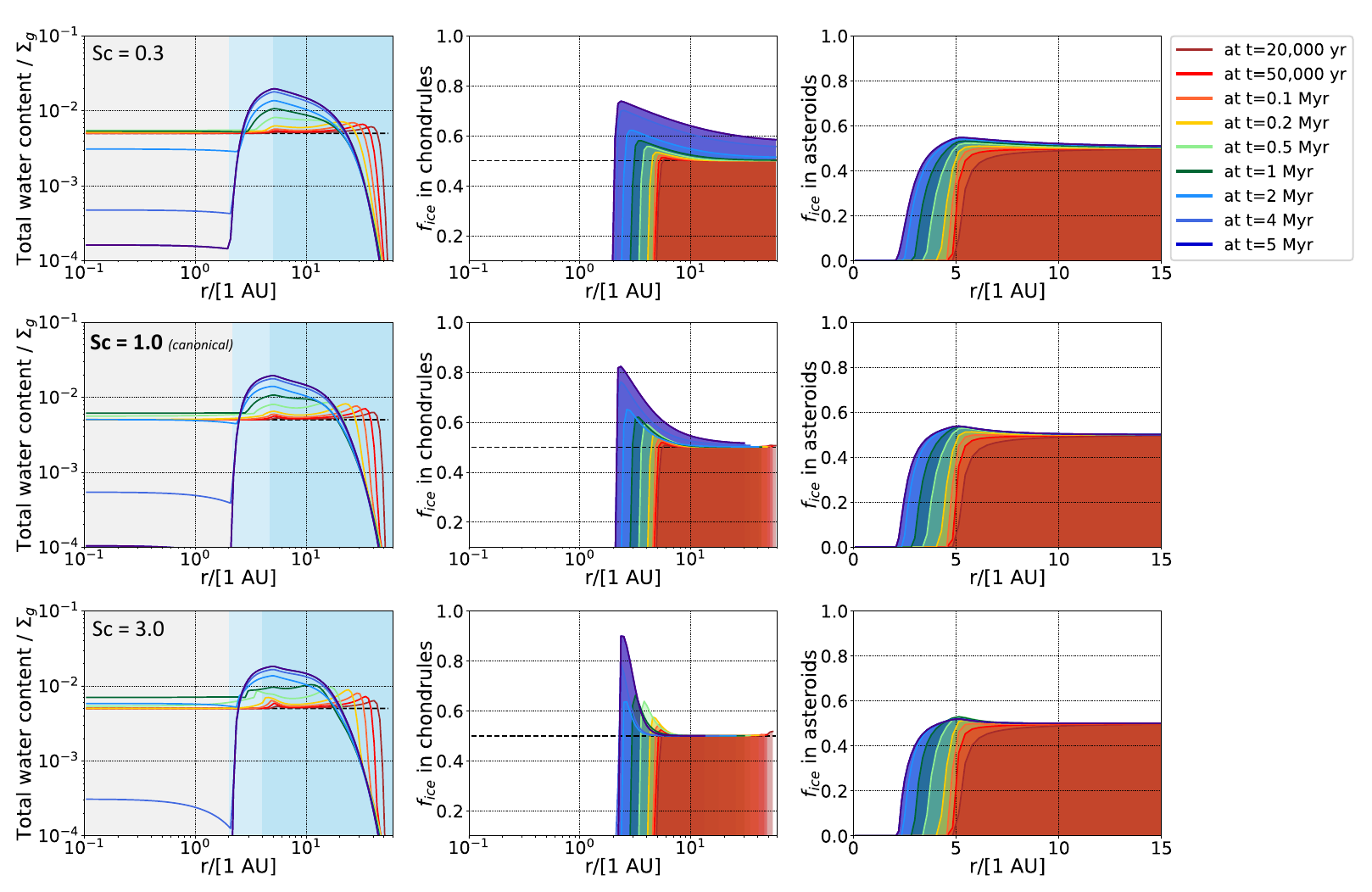}
\caption{Plots show results of variation of parameter Schmidt number (Sc) in an uniform $\alpha$ disk. Higher Sc (lower diffusivity) leads to higher peak abundance of ice in chondrules beyond the snow line, with a narrow width. Lower Sc leads to lower peak abundance but with a broader width, dependent on the diffusion of icy chondrules in the outer nebula. Figures and colors same as previous plots.}
\label{paramsc}
\end{figure*}

To more closely examine the importance of changing the diffusivity vs. drift of particles, we explored a range of Schmidt
number, from ${\rm Sc} = 0.3$, to 1 (our canonical case), to 3. The particle diffusivity is 
${\cal D}_{\rm p} = \nu / {\rm Sc}$, so higher Schmidt number yields lower particle diffusivity, but does not affect 
the drift rate or evolution of the disk.
Results are plotted in Figure \ref{paramsc}.
The cases with ${\rm Sc} = 3$ have the lowest diffusivity and therefore the highest relative rate of particle drift to 
particle diffusion. As in the case with $a = 0.06$ cm, which also had a high relative rate of drift to particle diffusion,
this leads to a higher ice fraction of chondrules beyond the snow line.
Comparing the different Schmidt number cases, those with the highest ${\rm Sc}$ are marked by the narrowest radial 
extent of enhanced chondrule ice fraction, and those with the lowest ${\rm Sc}$ have the broadest radial extent of 
high water fraction.
This is understood in terms of icy chondrules diffusing further 'upstream' beyond the snow line, widening the distribution,
when ${\rm Sc}$ is low and their diffusivity high. 
The water vapor abundance in the inner disk is also slightly lower for low Schmidt number, as chondrules can diffuse
back across the snow line more rapidly, and can more effectively capture water vapor beyond the snow line. 

\subsubsection{Effect of varying asteroid growth timescale}

\begin{figure*}
\plotone{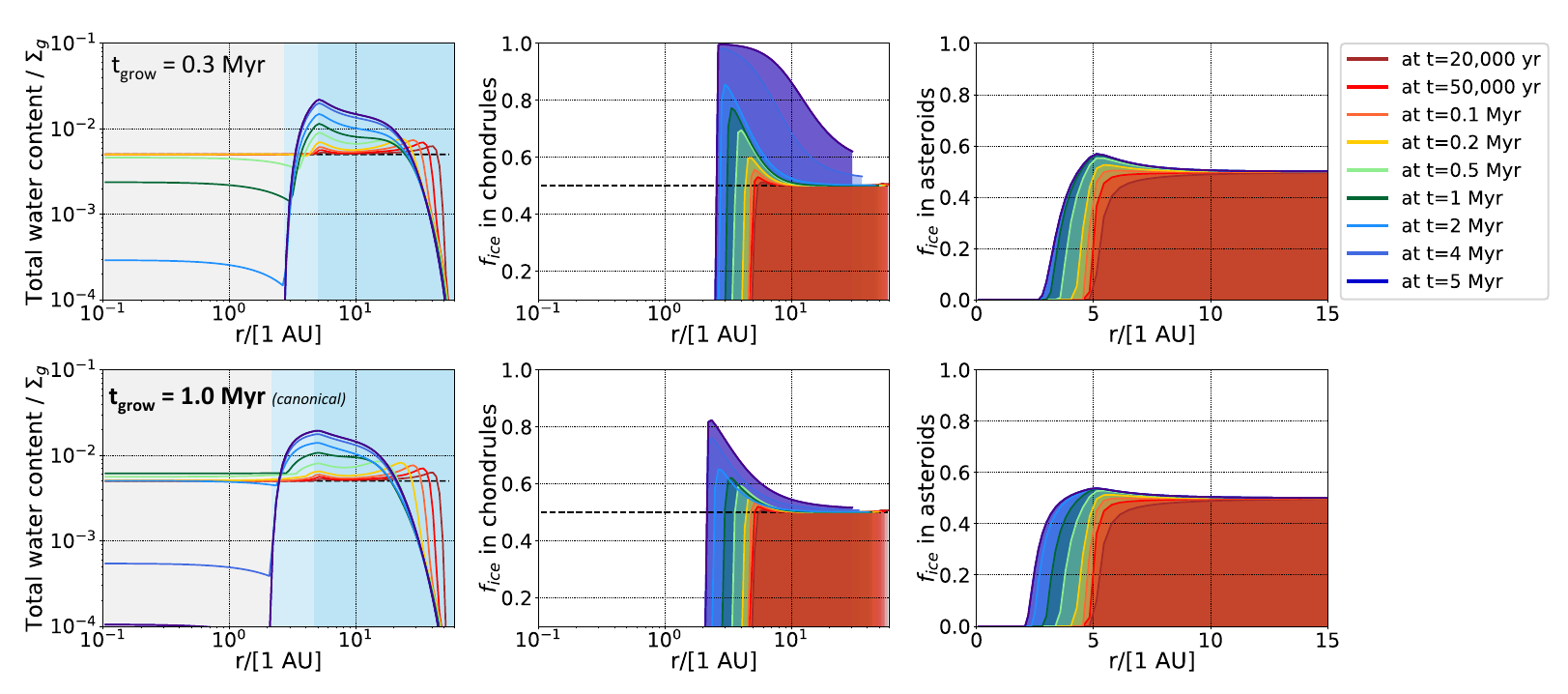}
\caption{Plots show results of variation of parameter growth timescale of asteroids $t_{\rm grow}$ in an uniform $\alpha$ disk. More rapid growth of asteroids depletes the pool of chondrules for diffusion through the snow line. Figures and colors same as previous plots.}
\label{paramtgrowth}
\end{figure*}

We explored the effect of varying the timescale $t_{\rm grow}$ over which chondrules are converted into larger,
asteroid-sized bodies. 
Asteroids are assumed to grow by converting a fraction of the mass (both silicate and ice) present at a given radius
into large bodies that from that point on are immovable. 
Beside our canonical case of $t_{\rm grow} = 1$ Myr, we consider $t_{\rm grow} = 0.3$ Myr, basically assuming that
planetesimals form three times faster than in our canonical case. 
The results are plotted in Figure \ref{paramtgrowth}.
Comparing the $t_{\rm growth} = 0.3$ Myr case to our canonical case, 
chondrules are more efficiently converted into asteroids before they can drift or diffuse. 
This significantly reduces $\Sigma_{\rm chon}$, both inside and outside the snow line. 
A dip in the water vapor abundance just inside the snow line (Figure \ref{paramtgrowth}a) reveals that inward transport of ice 
(through diffusion and also drift) is not fast enough to replenish water vapor, which continues to diffuse
outward across the snow line.
This results in an enhanced ice-to-rock ratio beyond the snow line.
These effects transport water outward so effectively, the snow line is moved slightly outward.

\subsubsection{Effect of varying the opacity}

\begin{figure*}
\plotone{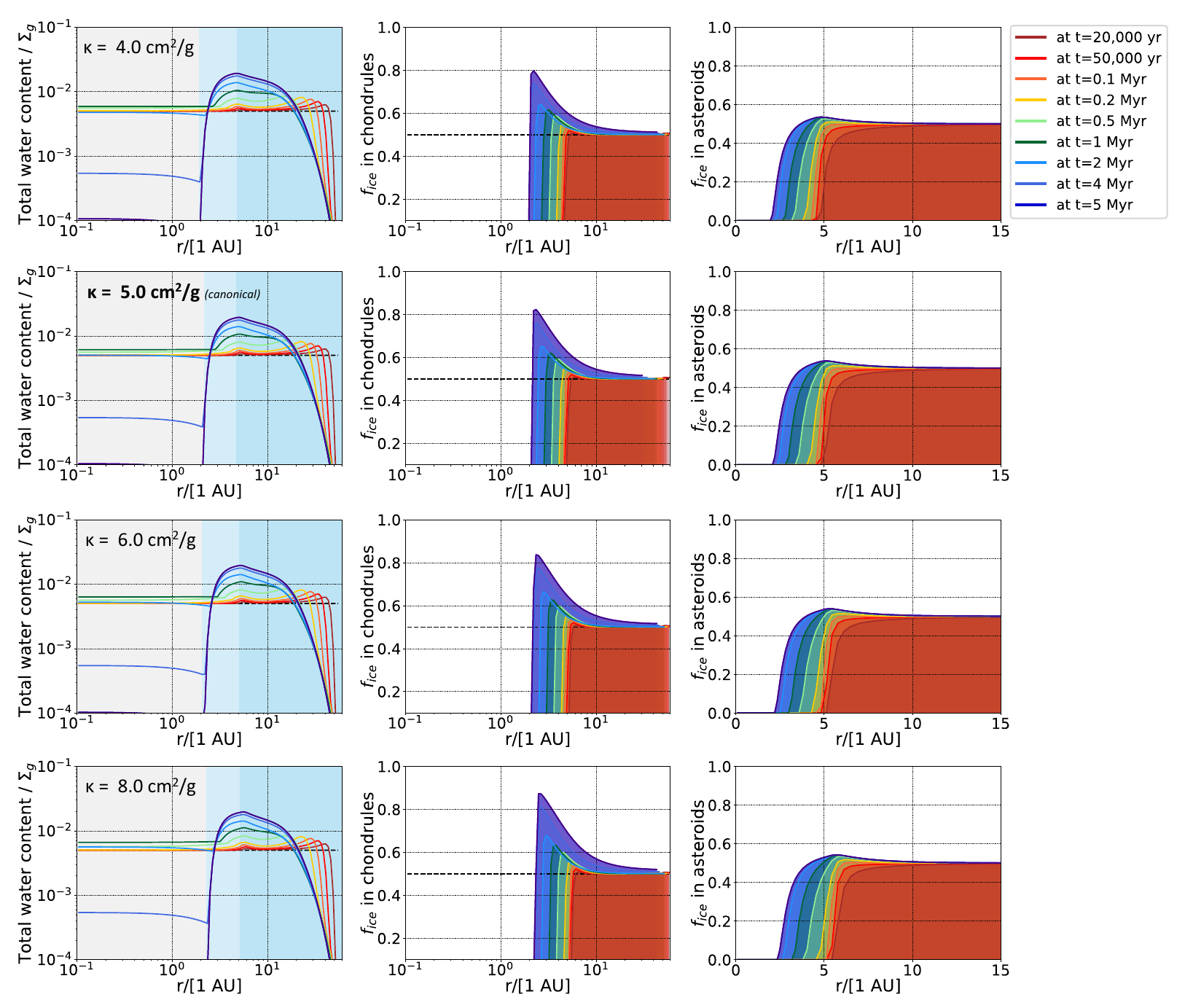}
\caption{Plots show results of variation of parameter $\kappa$ in an uniform $\alpha$ disk. Higher $\kappa$ leads to a warmer disk, that enhances both diffusion and drift rates to a similar extent. Diffusion is still predominant; therefore higher $\kappa$ therefore leads to a slight increase in peak chondrule ice abundance beyond the snow line. Figures and colors same as previous plots.}
\label{paramkappa}
\end{figure*}

We explored the effect of varying the opacity of the gas (due to dust), from $\kappa = 4 \, {\rm cm}^{2} \, {\rm g}^{-1}$,
to $\kappa = 5 \, {\rm cm}^{2} \, {\rm g}^{-1}$ (our canonical case), to $\kappa = 6 \, {\rm cm}^{2} \, {\rm g}^{-1}$, 
to $\kappa = 8 \, {\rm cm}^{2} \, {\rm g}^{-1}$.
Increased opacity leads to higher disk temperatures in actively accreting disks. 
Far from the star, the disk temperature conforms to the passively heated disk limit, but the snow line location 
is usually located where accretion has increased the disk temperature above the passively heated disk case.
Therefore higher opacities can increase disk temperatures and move the snow line outward.
Figure \ref{paramkappa} plots the effects on the radial distribution of water, which turn out to be minimal.
The location of the snow line hardly moves.
As Equation 2 demonstrates, the temperature in the disk varies as $\kappa^{1/3}$. 
As the temperatures around the snow line tend to fall as $r^{-1}$, the location of the snow line 
tends to vary as $r_{\rm snow} \propto \kappa^{-1/3}$. Across the entire range of opacities we considered, 
the location of the snow line varies by only 25\%, from about 2 AU to 2.5 AU. 

Increases in global disk temperatures slightly increase the turbulent viscosity, since 
$\nu = \alpha C H = \alpha (k T / \bar{m}) / \Omega$. This does not change the viscosity of gas or 
diffusivity of particles at the snow line, which is at more-or-less fixed temperature $\approx 160$ K.
The diffusion and drift of vapor and particles across the snow line would not differ from the canonical case.
But global increases in temperature and $\nu$ would decrease the disk evolution timescale $\propto r^2 / \nu$,
making the disk evolve slightly faster ($\approx 25\%$).
At any instant in time this would lead to greater ice-to-rock ratios of chondrules beyond the snow line. 
Overall the radial distribution of water ice (especially in asteroids) is little changed.

\subsubsection{Summary}

\begin{figure*}
\plotone{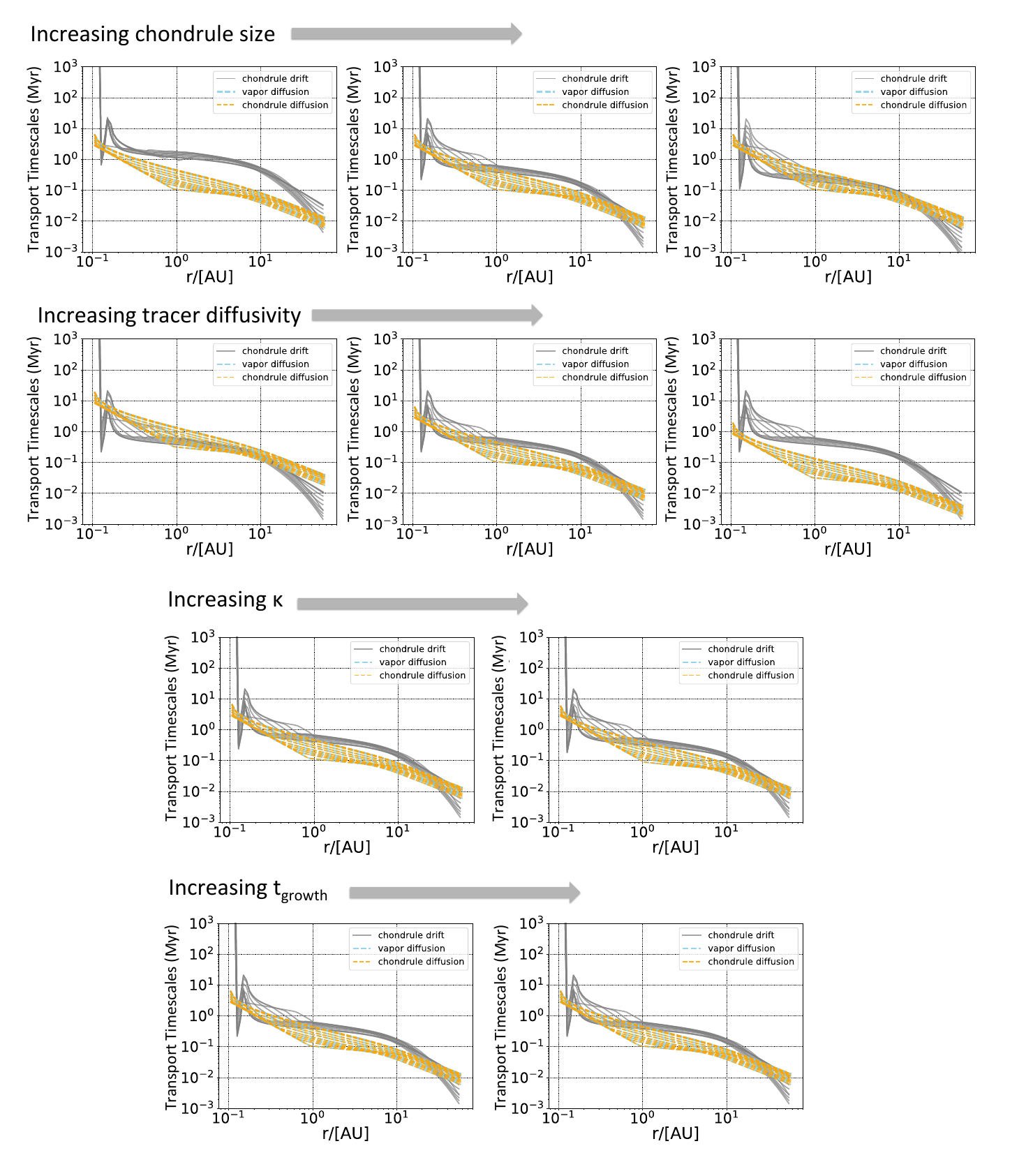}
\caption{Plots show relative timescales of the different radial transport processes for all the simulations performed in our parameter study, that include varying i) size of the chondrule; ii) diffusivity of water vapor; iii) timescale for growth of chondrules to asteroids; and finally iv) opacity of the disk material. While opacity has little effect on water distribution, $a_{\rm chon}$ and $t_{\rm growth}$ matter most, significantly affecting the abundance of water ice beyond the snow line.}
\label{alltimescales}
\end{figure*}

The cases considered above show the effects of varying chondrule size and Schmidt number, as well as asteroid 
growth timescale and disk opacity. 
Disk opacity is found to have little effect on the radial distribution of water, and especially the final water
content of asteroids.
One of the most important parameters affecting the distribution of water is the timescale on which chondrules
are converted into asteroids. 
For shorter timescales $t_{\rm growth}$, chondrules are more depleted from beyond the snow line and can
carry less water inward across the snow line by drift. This enhances the water-to-rock ratio. 
The other two parameters ($a$ and Sc) are of moderate importance, and highlight the subtle interplay between drift and diffusion
of vapor and particles. 
We plot in Figure \ref{alltimescales} the timescales of chondrule drift and diffusion and vapor diffusion in these different model disks. 
As previously seen, the cases where particle drift timescales are shorter than the chondrule diffusion and vapor 
diffusion timescales in the 1-10 AU region are the cases that lead to the highest water-to-rock ratios beyond the 
snow line.
This is somewhat unexpected, as faster-drifting chondrules might be expected to carry ice out of this region,
into the inner disk, faster than water vapor could diffuse back across the snow line. 
We attribute the enhancement in water-to-rock ratio to an overall increase in the water being brought to the snow line 
region from far out in the disk by drifting chondrules.

\begin{figure*}
\plotone{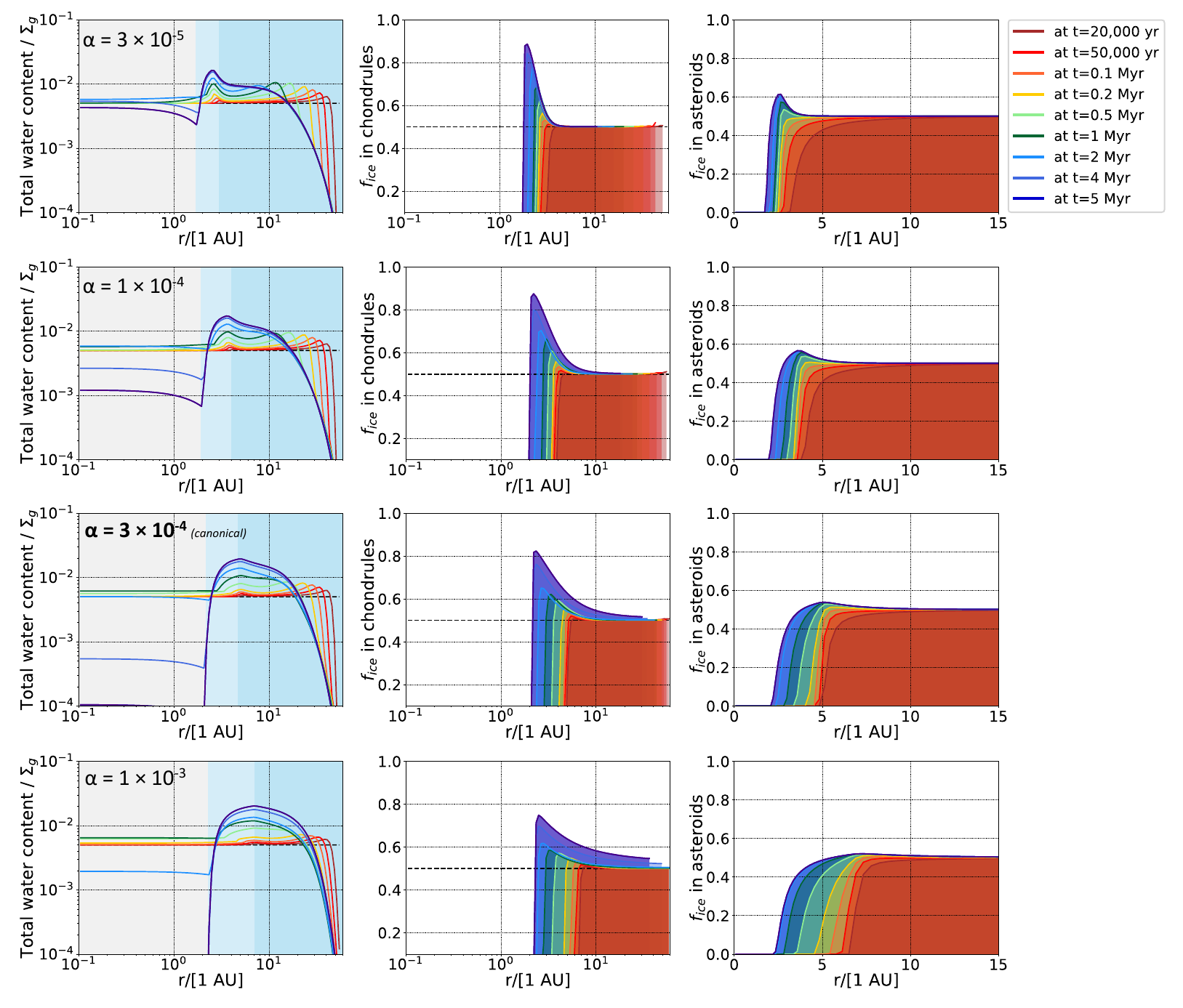}
\caption{Plots show the results of the different uniform $\alpha$ cases that employ a range of the globally uniform $\alpha$ as indicated. (a) shows total water content/$\Sigma_{g}$ with $r$, where total water = $\Sigma_{\rm vap}+\Sigma_{\rm chon}+\Sigma_{\rm ast}$; (b) shows radial variation in the water-rock ratio in chondrules; (c) shows radial variation in the water-rock ratio in asteroids that grow from chondrules. Figures and colors same as previous plots.}
\label{differentunialpha}
\end{figure*}

Having examined the roles of different physical processes in the disk, we 
next explore the effect of varying the uniform value of $\alpha$ across the disk, considering cases
ranging from $\alpha = 3 \times 10^{-5}$, to $\alpha = 1 \times 10^{-4}$, to $\alpha = 3 \times 10^{-4}$ 
(our canonical case), to $\alpha = 3 \times 10^{-3}$. 
The results are plotted in Figure \ref{differentunialpha}.
With increasing values of $\alpha$, the inner disk within the snow line is increasingly depleted in water vapor. 
At 5 Myr, the water abundance declines from $4.5 \times 10^{-3}$ for $\alpha = 3 \times 10^{-5}$, 
to $\ll 10^{-4}$ for $\alpha = 1 \times 10^{-3}$.
As $\alpha$ is increased, water vapor back-diffuses outward across the snow line more quickly, but water ice
on particles also diffuses inward more quickly.
Increasing $\alpha$ also would increase the inward velocity of gas, $V_{{\rm g},r}$, but from 
Equation 9, as long as ${\rm St} \ll 1$, the drift speed of chondrules is little affected by changes in $\alpha$.
Therefore increasing $\alpha$ increases the relative importance of particle diffusion to drift. 
This conclusion is supported by Figure \ref{alltimescales}, which shows that drift is more rapid at low values of $\alpha$, 
but for high values of $\alpha$ diffusion is more rapid than drift. 
At higher $\alpha$, icy chondrules beyond the snow line are better able to diffuse outward, giving them more 
opportunity to be incorporated into asteroidal material.
For all values of $\alpha$, there is a zone beyond the snow line in which the ice-to-rock fraction is increased. 
As $\alpha$ increases, the width of that zone increases, but the peak ice-to-rock concentration decreases. 
This trend is somewhat echoed in the plots of asteroid water-ice fraction plots (third column of Figure \ref{differentunialpha}).
Higher values of $\alpha$ yield a broader range of radii over which asteroid water fractions increase from
0wt\% to 50wt\%. Smaller values of $\alpha$ allow for slightly higher water-ice fractions, but the effect is muted. 
In summary, higher values of $\alpha$ yield higher accumulations of ice beyond the snow line, but do not 
yield the highest peak abundances of water ice beyond the snow line, as the water content beyond the snow line 
is distributed over a broader radial region.

\begin{figure*}
\plotone{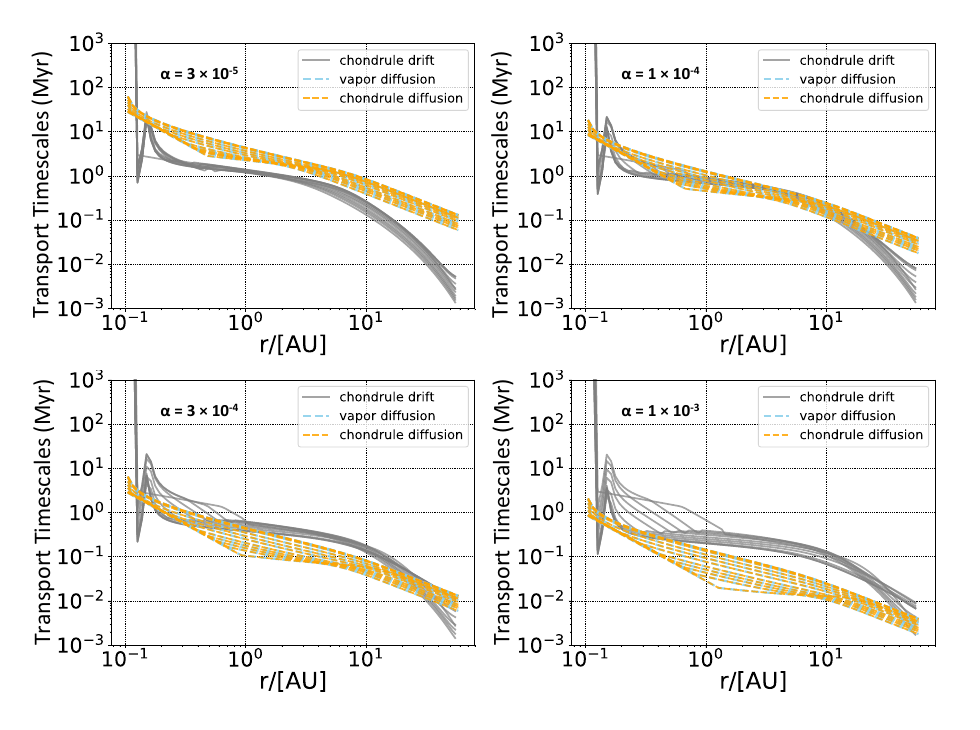}
\caption{Plots show the timescales of the various radial transport processes of volatiles for different uniform $\alpha$ cases (including the canonical case of uniform $\alpha$). Colors and lines same as Figure \ref{unialphatimescale}. Note that diffusion becomes quicker than drift and therefore is more important as $\alpha$ is increased from $3 \times 10^{-5}$ (upper left plot) to $1 \times 10^{-3}$ (bottom right plot).}
\label{difftimescalesunialpha}
\end{figure*}
\subsection{Water distribution with an MRI $\alpha$ profile}

\begin{figure*}
\plotone{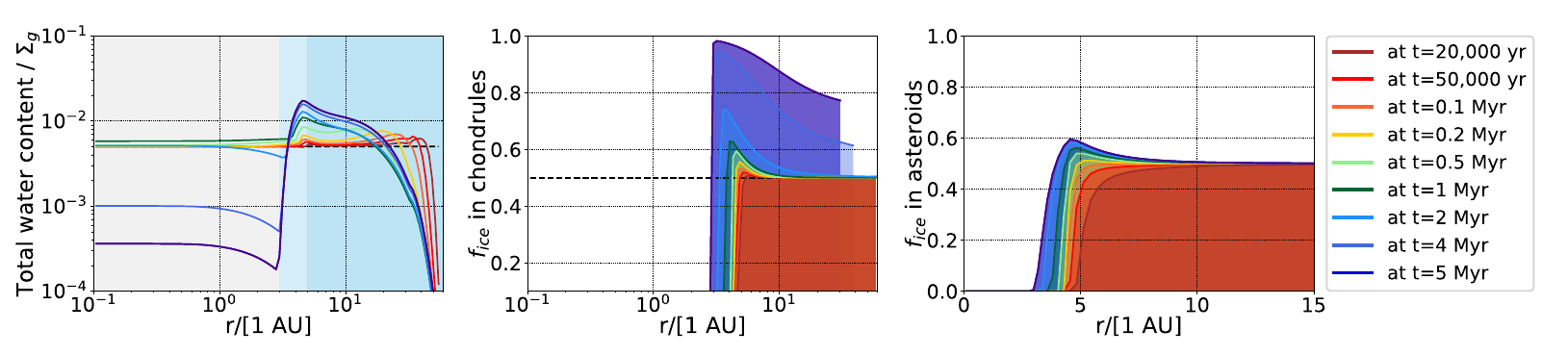}
\caption{Plots show results of the MRI-$\alpha$ profile, i.e., CASE II. (a) shows total water content/$\Sigma_{g}$ with $r$, where total water = $\Sigma_{\rm vap}+\Sigma_{\rm chon}+\Sigma_{\rm ast}$); (b) shows radial variation in the water-rock ratio in chondrules; (c) shows radial variation in the water-rock ratio in asteroids that grow from these chondrules. See Figures \ref{sigmaprof} and \ref{tempprof} for corresponding surface density and temperature radial profile plots. Colors and lines same as in Figure \ref{unialpha}.}
\label{mrialpha}
\end{figure*}

Case I examines the distribution of water in a disk with uniform $\alpha = 3 \times 10^{-4}$.
Case II, presented in Figure \ref{mrialpha}, shows the distribution of water in a disk subject to turbulent viscosity
like that driven by the MRI. 
An immediate result is that the ice-to-rock fraction of chondrules beyond the snow line is much higher than
in Case I, with an ice fraction of 0.97. 

The difference in $\alpha$ at the snow line is not large, less than a factor of 2. 
Larger differences in $\alpha$ than this did not lead to such large differences in water content (\S 3.1).
We attribute the increased water-ice fraction to several factors. 
One is the surface density, $\Sigma(r)$, which we plot in Figure \ref{sigmaprof}. 
The effect on the temperature is seen in Figure \ref{tempprof}, and the timescales for diffusion and drift are
plotted in Figure \ref{unialphatimescale}. 
As seen in Figure \ref{sigmaprof}, by as early as 0.5 Myr, the surface density at 2 AU is almost a factor of 4 larger 
in Case II than in Case I. 
This is due to $\alpha$ being lower in this region than surrounding regions in Case II, causing mass to pile up
in this same region where the snow line is located.
The lower values of $\alpha$ in this region also lead to vapor and chondrules diffusing less rapidly there, 
but more rapidly in the surrounding regions. 
At the snow line itself vapor and icy chondrules do not diffuse as efficiently. This leads to the dip in the profile of the total water content to bulk gas just within the snow line (Figure \ref{mrialpha}a).
At the same time, the higher densities lead to lower Stokes numbers, which leads to smaller drift speeds. Moreover, the snow line does not move as much as in the uniform $\alpha$ case. This retains a higher concentration of ice-rock over a smaller region than in the uniform $\alpha$ case. Over and above this, the icy chondrules quickly diffuse outward into the outer disk where they can grow and be accreted in asteroids and stay there, without the risk of the ice being lost as vapor near the snow line. 
All of these factors combine to bring water ice or vapor to the snow line region, but to inhibit its escape
from the snow line region. 

\begin{figure*}
\plotone{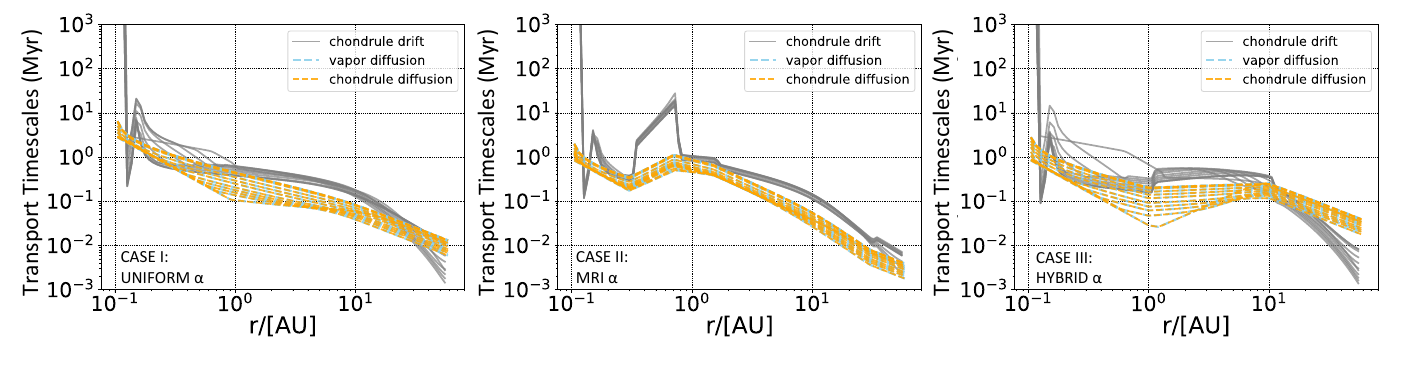}
\caption{Plots show the timescales of the various radial transport processes for volatiles for the 3 $\alpha(r)$ profiles considered in this work;  i) the Uniform $\alpha$ profile, ii) MRI-$\alpha$ profile, and iii) Hybrid $\alpha$ profile. Colors same as in Figure \ref{unialphatimescale}.}
\label{timescaleofthethreecases}
\end{figure*}

\begin{figure*}
\plotone{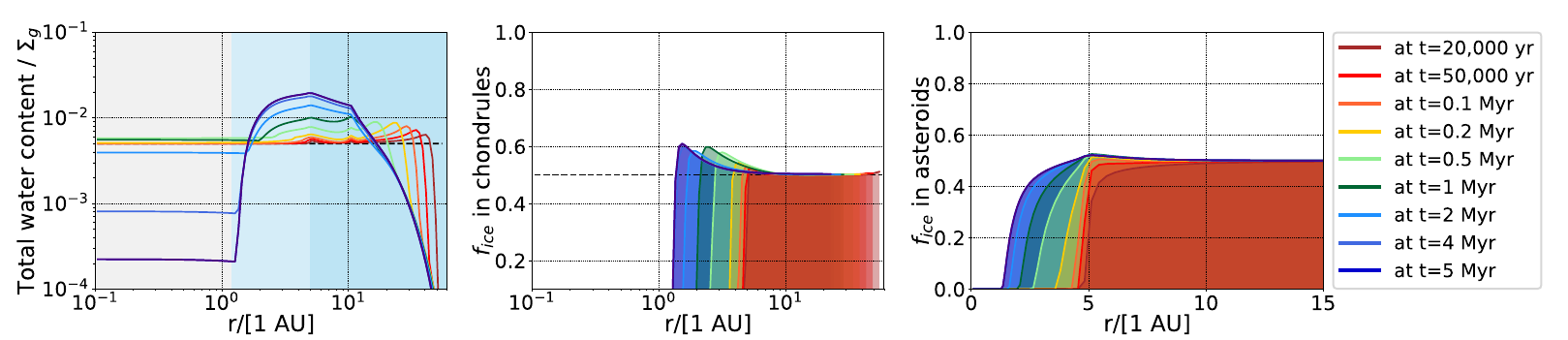}
\caption{Results of the hybrid-$\alpha$ profile, i.e., CASE III. (a) shows total water content/$\Sigma_{g}$ (total water = $\Sigma_{\rm vap}+\Sigma_{\rm chon}+\Sigma_{\rm ast}$) with $r$; (b) shows radial variation in the water-rock ratio in chondrules; (c) shows radial variation in the water-rock ratio in asteroids. See Figures \ref{sigmaprof} and \ref{tempprof} for corresponding surface density and temperature radial profile plots. Colors and lines same as in Figure \ref{unialpha}.}
\label{hybridalpha}
\end{figure*}

\subsection{Water distribution with the hybrid $\alpha$ profile}

Case III, presented in Figure \ref{hybridalpha}, shows the distribution of water in a disk subject to turbulent 
viscosity like that driven by VSI plus disk winds in the inner disk.
An important difference between this case and the others is that the surface density $\Sigma(r)$ is flat
and significantly lower than the other cases, with $\Sigma(r) \approx 500 \, {\rm g} \, {\rm cm}^{-2}$
throughout the snow line region by 0.5 Myr (Figure \ref{sigmaprof}).
The higher values of $\alpha$ in the inner disk keep temperatures in the inner disk hotter than in Cases I and II, 
and allows the inner disk to evolve and lose mass more rapidly.
Compared to the other cases, initially the temperatures are higher in the inner disk due to the higher $\alpha$,
but then as surface densities decrease, the temperatures drop more precipitously and the snow line moves in rapidly.
Between 0.02 and 5 Myr the snow line moves in from 5 AU to 1 AU as temperatures drop (Figure \ref{tempprof})

Rapid inward drift of the snow line keeps the peak abundance at about the same level throughout the simulation from
1 - 5 AU. 
This results in a large ``transition" region in the asteroid water-ice fraction (Figure \ref{hybridalpha} c) between the water ice 
solid material beyond 5 AU to water-depleted inner disk within the snow line. 
The vapor content is seen to deplete to around the same levels in the inner disk as the uniform $\alpha$ case.

\subsection{Convergence tests}

\begin{figure*}
\plotone{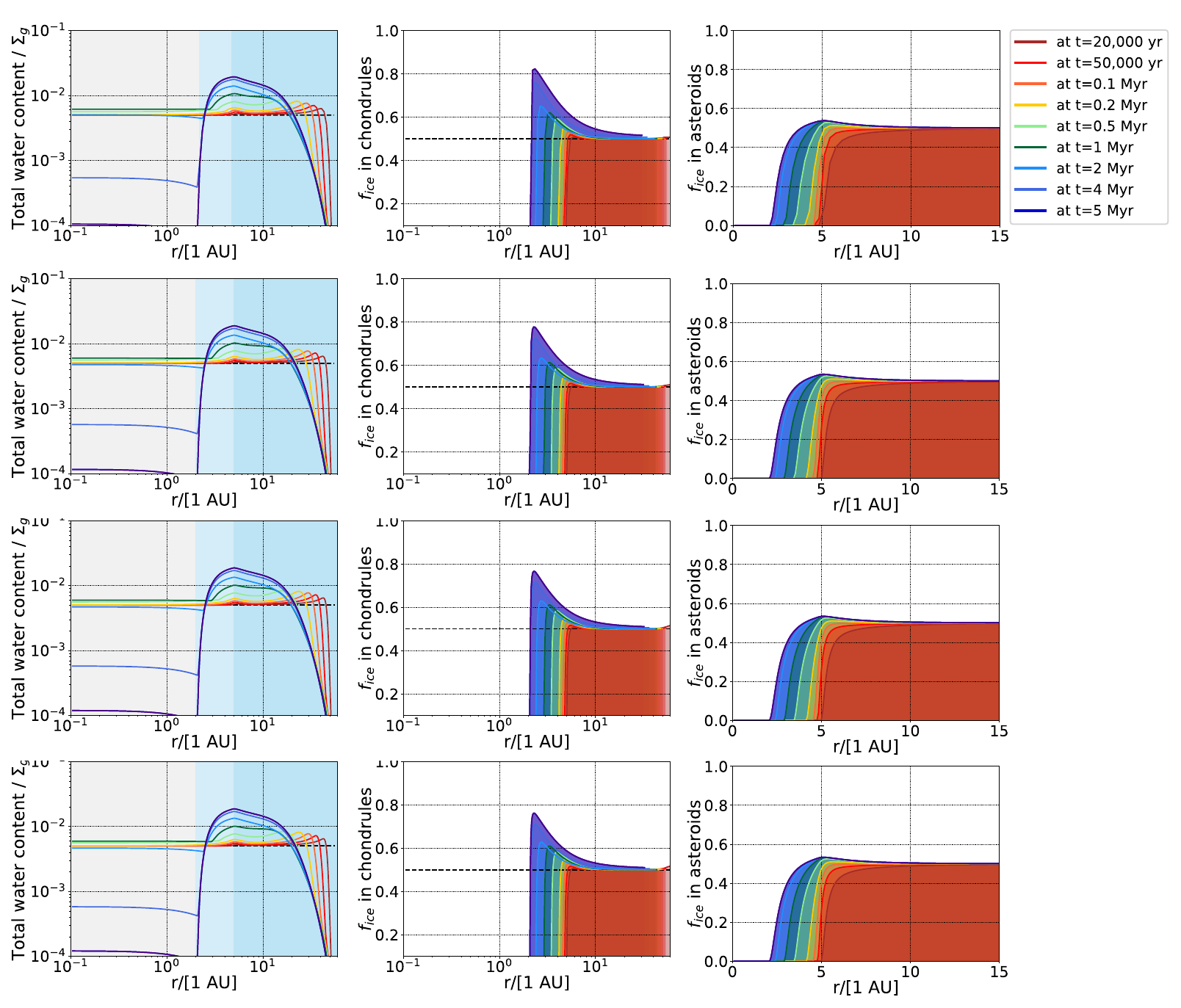}
\caption{Plots show the results of our convergence study. Resolution increases from top to bottom with number of radial zones increasing as 100, 200, 300 and 400 radial zones}
\label{convergence}
\end{figure*}

We also performed convergence tests to ensure that our simulations returned similar results with increase in 
our grid resolution. 
We performed the same simulation with 200, 300 and 400 zones, and found that with higher resolution, the peak 
water-to-rock ratio beyond the snow line dropped by less than 4\%, and that none of our results or conclusions
is significantly changed.
We find no other significant variation due to increase in resolution (Figure \ref{convergence}). 

\section{Discussion}

\subsection{Sensitivity Analysis} 

In our parameter study of uniform-$\alpha$ cases, we considered the effects of several parameters on the
distribution of water ice in the disk and in planetesimals that would eventually form planets. 
These included chondrule size, Schmidt number, the growth timescale of asteroids, the opacity, and the 
value of $\alpha$.
Our sensitivity analysis allowed us to identify which of these factors had the biggest impact on the 
distribution of water ice, both in chondrules and in the planetesimals formed from them. 

\subsubsection{Parameters affecting water distribution in the disk} 

We find that for variations across the likely range of each input, the size of particles, i.e., chondrule
radius $a$, had the largest effect on the water-ice fraction in chondrules (Figure \ref{paramachon}).
In our canonical case with $a = 0.03$ cm, the water ice fraction in asteroids just beyond the snow line at
5 Myr slightly exceeded 80\%. For smaller particles, $a = 0.01$ cm, the fraction barely exceeded 60\%, while
for slightly larger particles, $a = 0.06$ cm, the ice fraction approached 100\%. 
Changing the size of small particles onto which water vapor can condense as ice, by factors of just 2 to 3,
led to very large changes in the ice-to-rock ratio on chondrules.
The cases with larger chondrules also saw the water-ice fraction reach elevated levels at far greater radii 
beyond the snow line, 

The next most important parameter is the growth timescale of asteroids $t_{\rm grow}$ (Figure \ref{paramtgrowth}).
For our canonical case with $t_{\rm grow} = 1$ Myr, the water-ice fraction is $\approx 80\%$ beyond the snow line 
at 5 AU.
For a faster conversion rate of chondrules into planetesimals, $t_{\rm grow} = 0.3$ Myr, the water-ice fraction
approaches 100\%. Far fewer chondrules are available for the water ice to adhere to, leading to higher water-ice
fractions.
The width of the water-ice region beyond the snow line is slightly larger as well. 

Variations in the value of $\alpha$ were important but not as prominent as the changes in the above parameters
(Figure \ref{differentunialpha}). 
The water-ice fraction beyond the snow line at 5 Myr is $\approx 80\%$ in our canonical case with
$\alpha = 3 \times 10^{-4}$.
For lower values of $\alpha = 3 \times 10^{-5}$ this fraction approaches 90\%, while for larger values of 
$\alpha = 1 \times 10^{-3}$ it is 75\%. We also find that as $\alpha$ is increased, we find that vapor concentration decreases by just short of two orders of magnitude at t = 5 Myr between our lowest and highest adopted $\alpha$ (10$^{-5}$ to 10$^{-3}$) values. 

Variations in the diffusivity of small particles, via the Schmidt number ${\rm Sc}$, led to notable but smaller
changes in the water-ice fraction in chondrules ({\rm Figure \ref{paramsc}). 
Beyond the snow line at 5 AU, the water-ice fraction is $\approx 80\%$ in our canonical case with ${\rm Sc} = 1$,
dropping slightly to 75\% for ${\rm Sc} = 0.3$, and rising to 90\% for ${\rm Sc} = 3$. 
The width of the enhanced water-ice region is more strongly affected by the particle diffusivity, reaching
to much greater radii for greater diffusivity (lower ${\rm Sc}$).

Surprisingly, changes in the value of the opacity, $\kappa$, had the smallest effect on the distribution of water ice
in the disk. 
The cases with $\kappa = 4 \, {\rm cm}^{2} \, {\rm g}^{-1}$ and $\kappa = 8 \, {\rm cm}^{2} \, {\rm g}^{-1}$ showed
almost no difference in peak water-ice content or distribution of water ice beyond the snow line, compared to our 
canonical case with $\kappa = 5 \, {\rm cm}^{2} \, {\rm g}^{-1}$.
Opacity sets the temperature in the disk and the location of the snow line, but it is not a large effect. 
From Equation 2, the temperature in actively accreting disks is $T \propto \kappa^{1/3}$, and from 
Figure \ref{tempprof} temperature also is dropping roughly as $T \propto r^{-1}$ in the snow line region, meaning that
the location of the snow line varies roughly as $\kappa^{1/3}$, being only tens of percent closer in or farther
out across this range of opacity. 
More importantly, the distribution of ice and vapor is little affected by these changes. 

We conclude that ``planetary" parameters like the radius of small particles (chondrules) or the rate at which they
are converted into planetesimals, dominate the distribution of water ice in the disk, and not such ``disk" parameters
like $\alpha$ or opacity $\kappa$. 

\subsubsection{Parameters affecting distribution of water ice in planetesimals}

The distribution of water ice in a disk translates into water content of planets only by affecting the water 
content of asteroidal planetesimals, which are the building blocks of planets.
Planetesimals, we assume, grow continuously over the lifetime of the disk, converting small chondrules into 
large asteroids with an $e$-folding time of $t_{\rm grow}$.
Because asteroids grow over a long timescale, instantaneous distributions of water ice in the disk are 
averaged out. Most of the cases considered above lead to similar distributions of water ice in planetesimals.
Peak water-ice contents rarely exceed 50\%, sometimes approaching 60\%. 
Beyond the snow line, almost all planetesimals in all cases have $\approx 50\%$ ice. 
The quantity that is strongly affected is the radial distribution of asteroids with intermediate water ice 
content $< 50\%$. 

In our canonical case, that distribution at 1 Myr extends from about 2.5 AU, where water ice contents are barely
above 0\%, to about 5 AU, where they approach 50\%.
Most parameters have little effect on this distribution: this same pattern is observed across the range we consider 
for chondrule size $a$, Schmidt number ${\rm Sc}$, asteroid growth timescale $t_{\rm grow}$, and opacity $\kappa$
(Figures \ref{paramachon}, \ref{paramsc}, \ref{paramtgrowth}, and \ref{paramkappa}).

The radial distribution of asteroid water-ice content seems most affected by $\alpha$ (Figure \ref{differentunialpha}). 
The largest values of $\alpha \sim 10^{-3}$ allow the disk to evolve more rapidly; asteroids less reflect a
snapshot of the disk and more reflect an average over time.
The peak water ice concentrations do not exceed 50\% anywhere.
For the smallest values of $\alpha = 3 \times 10^{-5}$, the asteroids beyond the snow line can reach
60\% ice, with a drop-off beyond that. 
To a lesser extent, the asteroid growth timescale matters, with $t_{\rm growth} = 0.3$ Myr yielding 
asteroids beyond the snow line with $>55\%$ water ice.
The existence of a local maximum in the asteroid water ice content appears sensitive to the 
timescale of planetesimal formation ($t_{\rm growth}$) relative to the disk evolution timescale 
($\sim r^2 / \nu \propto \alpha^{-1}$). 
Faster asteroid growth and/or slower disk evolution allows a local maximum.

We conclude that water ice content of planetary materials is affected equally by ``disk" parameters 
($\alpha$) and ``planetary" parameters ($t_{\rm growth}$). 

The sensitivity of the water ice distribution in the protoplanetary disk to various input parameters
can be understood largely in terms of the relative rates of diffusion and drift.
As seen in Figures \ref{alltimescales}, \ref{difftimescalesunialpha} and \ref{timescaleofthethreecases}, in general diffusive transport of vapor and chondrules is
faster than drift of chondrules, by almost an order of magnitude throughout much of the disk in most of the 
presented simulations.
In the common situation where drift is much slower than diffusion, the water ice content just beyond the snow line 
is close to the cosmic abundance of water ice, with a water-ice fraction of 50\%. 
But as the rate of drift is comparable to, or faster than, the rate of diffusion, then the water-ice fraction
in solids beyond the snow line can be become quite large.

\subsection{Physical processes affecting water distribution}

The radial distribution of water ice in chondrules beyond the snow line can be understood in part as a 
relative rate of particle drift vs. diffusion of particles and vapor. 

For example, Figure \ref{differentunialpha} shows that the peak water-ice fraction is higher at 5 Myr ($\approx 90\%$) for 
the case with $\alpha = 3 \times 10^{-5}$ than it is ($\approx 75\%$) at 5 Myr in the disk with $\alpha = 1 \times 10^{-3}$. 
This is not due to temporal differences: these water-ice enhancements occur late in disk evolution, so if anything
the disk with higher $\alpha$ should develop the same water-ice enhancements at earlier times. 
(The evolution timescale scales as $r^2 / \nu \propto \alpha^{-1}$)

It is also not due to a relatively higher general rates of diffusion. Disks with large $\alpha$ should cause water
vapor to diffuse outward past the snow line faster, leading to greater enhancements; or, including the inward diffusion
of icy chondrules through the snow line, one might expect the two effects to cancel each other out, leading to 
no differences. 
Figure \ref{difftimescalesunialpha} demonstrates that the diffusion timescales of vapor and particles at the snow line vary from $\sim 2$ Myr 
for $\alpha = 3 \times 10^{-5}$, to $\sim 0.02$ Myr for $\alpha = 1 \times 10^{-3}$, without a significant change 
in the water-ice fraction beyond the snow line. 

A change in just the diffusion rate of particles but not vapor leads to somewhat larger changes. 
Figure \ref{paramsc} demonstrates that as the diffusion rate of particles alone is varied by an order of magnitude,
as ${\rm Sc}$ is varied from 0.3 to 3, the water-ice fraction does change, but only from about 75\% to 90\%.
With increased ${\rm Sc}$, particle diffusivity decreases, meaning that icy chondrules diffuse inward more slowly,
leading to a lower loss of icy solids from the region beyond the snow line. 
Particle diffusion is not the dominant loss mechanism, however. 

Rather, the dominant effect appears to be particle drift, and factors affecting drift rates appear to be the
main determinant of the water-ice distribution.
Figure \ref{paramachon} shows that as chondrule radii are increased from 0.01 cm to 0.06 cm, there is very little change
in the diffusion timescale. 
This is because ${\cal D} = \nu \, (1 + {\rm St}^2)^{-1}$, and ${\rm St} \ll 1$ for chondrules of all these sizes.
The drift rate changes considerably, though, as the drift velocity is proportional to ${\rm St}$ and therefore
$a / \Sigma$. The large chondrules drift 6 times faster than the small chondrules, and the drift timescales 
reflect this. 
Larger drift rates might be expected to more quickly drain icy chondrules out of the region beyond the snow line, 
lowering the water-ice fraction there, opposite to the observed trend ; but a more important factor is the influx
of ice to this region from farther locales in the disk.
The drift velocity of particles is $V_{\rm drift} \propto {\rm St} \eta V_{\rm K}$ $\propto \eta V_{\rm K} (a / \Sigma)$.
The mass flux of particles is therefore $\Sigma V_{\rm drift}$ $\propto \eta V_{\rm K} a$.
The influx of icy particles into the region beyond the snow line therefore is 6 times larger for the 0.06 cm 
radius chondrules than for the 0.01 cm radius chondrules. 
This ice ultimately vaporizes inside the snow line, but as long as as significant fraction is returned to beyond
the snow line, the water-ice fraction there will ultimately be higher.

A re-examination of Figures \ref{differentunialpha} and \ref{difftimescalesunialpha} suggests that the cause of the increased water-ice fraction in 
cases with low $\alpha$ is primarily due to differences in drift rate. 
Because disks with low $\alpha$ evolve more slowly, they maintain high surface densities for longer.
Higher $\Sigma$ leads to slower drift rates.  Indeed, the cases with low $\alpha$ have the longest drift timescales,
and and the cases with highest $\alpha$ have the shortest drift timescales. 
The short drift timescales lead to more overall water ice beyond the snow line for the high-$\alpha$ cases,
as seen in the first column of Figure \ref{difftimescalesunialpha}.
The high $\alpha$ additionally has the effect of broadening and lessening the peak water-ice fraction. 

These overall trends help explain why factors like the disk opacity have little effect on the water-ice content.
Changes in opacity have a relatively minor effect on the temperature, which is proportional to $\kappa^{1/3}$. 
Factor of 2 increases in $\kappa$ have a minor effect on the disk viscosity, increasing it only by a factor of 1.26.
This would affect the diffusion of both particles and vapor, which would not directly change the distribution of
ice much. The surface density at any time would be slightly decreased by the higher viscosity, leading to higher
drift of particles that slightly increases the water-ice fraction. But because the effects are indirect, the 
change in the water-ice fraction is small.  

In this work, we don't consider the growth of the opacity-yielding fine dust grains into larger particles or into chondrules. We assume that they are created in collisions between chondrules and don't track water ice deposited on them. We argue that these assumptions are valid as our simulations reveal that the timescales of diffusion are smaller by an order of magnitude than drift, and chondrule diffusion (unlike drift) is almost independent of grain-size for the smallest sizes of dust grains (where St $<<$ 1). As long as the rates of grain growth by coagulation and fragmentation balance each other, and the collisional size distribution stays constant across the inner disk where the water snow line is located (independent of the power law of the size distribution, e.g., Brauer et al.\ 2008; Birnstiel et al.\ 2011), our model remains valid as diffusion is faster than drift. We find that our assumption of collisional equilibrium across the inner disk is consistent with Birnstiel et al.\ (2015; 2011), who argue for efficient fragmentation in the inner disk. Then, the smaller dust grains only carry a fraction of the total water mass but the trends of total water abundance and ice fraction in smaller dust grains across and beyond the snow line remain the same as with chondules. On the contrary, if chondrules beyond the snow line requires higher relative velocities for fragmentation, more water is carried by the largest particles which upon drifting inward more rapidly than smaller particles yield more vapor that is back diffused into the snow line, leading finally to more ice accumulation beyond the snow line.

Ultimately, the water-ice content of small solids beyond the snow line is set by subtle interplays between 
viscosity, diffusion, and drift, with factors affecting drift playing the most important role. 

\subsection{Effects of changing $\alpha(r)$}

With these insights, we can begin to understand the water distributions in our three different $\alpha(r)$ profiles.
Changing $\alpha(r)$ in the disk, either by varying a globally uniform value, or by adopting a radially varying 
$\alpha(r)$ profile, changes the distribution of ice by altering the diffusion rates of particles and vapor, and by
changing the drift rates of particles.

Figure \ref{hybridalpha} shows that Case III, with the hybrid $\alpha(r)$ profile due to hydrodynamic instabilities and 
and magnetic disk winds, and Case I, have the highest overall amounts of water beyond the snow line, with water fractions
reaching 4 times the overall disk mass fraction of $5 \times 10^{-3}$. 
Case II has water abundances approaching these levels, 3 times the background levels, but over a narrower range of radii. 

Figure \ref{mrialpha} shows that Case II, with the $\alpha(r)$ profile resembling that due to the MRI, has the highest 
local concentrations of water-ice beyond the snow line. 
Some regions show water-ice fractions of solids $> 95\%$, in contrast to the peak water-ice fractions 
$\approx 60\%$ in Case III. Case I with uniform $\alpha$ has intermediate peak values at 5 Myr of $\approx 80\%$.
Since all three cases have similar overall water contents, we attribute these effects largely to the different 
values of $\alpha$ at the snow line, which tends to smear out and decrease the peak distributions. 
At 5 Myr, the value of $\alpha$ at the snow line is only $1.7 \times 10^{-4}$ in Case II, allowing for sharp 
distributions. 
At the snow line at 5 Myr, $\alpha = 3 \times 10^{-4}$ in Case I, and is as large as 
$8 \times 10^{-4}$ in Case III. 

The different cases yield different positions of the snow line over time. 
In Case I, the snow line starts at about 4.5 AU and moves in to about 2 AU by 5 Myr.
In Case II, the snow line starts at about 5 AU and moves in to 3 AU, and for 
Case III it moves in from 5 AU to about 1.2 AU by 5 Myr. 
Case III shows the snow line moving in the most because the higher values of $\alpha$ at small radii
deplete the inner disk of gas the fastest, leading to decreased accretional heating. 

An important side effect of the snow line moving in so much more in Case III is that it leads to a greater 
diversity of asteroid water-ice fractions in the inner disk.
In Case III, asteroids forming at 2.5 AU between 0.5 Myr and 5 Myr might have anywhere from 0\% to 50\% ice, 
with similar ranges seen for asteroids forming anywhere between 2 and 4 AU. 
In contrast, in Case II, asteroid water content only varies in a relatively narrower range of radii, from
3 to 4 AU.
Case I, with uniform $\alpha$, is somewhat intermediate between these cases, with asteroids showing variable 
water content between 2.5 and 3.5 AU. 
These results in particular suggest ways to use the water distributions of asteroids to probe the 
$\alpha(r)$ profile of the solar nebula. 

In this study, we do not account for the temporal variation in $\alpha(r)$ and merely assume that it is a function of location in the disk. In the cases of MRI- and wind-driven disks, as $\Sigma$ changes with time, $\alpha$ is likely to change as well (eg. Kalyaan et al.\ 2015; Suzuki et al.\ 2016). While this is important to consider in the future, we argue that the focus of this paper is to understand the effect of the radial variation of $\alpha$ on the radial distribution of water. As $\alpha(r)$ itself has subtle effects on the various interconnected physical processes at the snow line, the effect of addition of $\alpha(t)$ at each $r$ adds another layer of complexity with subtle effects that can be more difficult to separate.

\subsection{Observational Tests} 

\subsubsection{Astronomical Observations of Disks}

Our work highlights the connections between different $\alpha(r)$ profiles and the distribution of water vapor 
in the inner disk or water-ice on chondrules or in asteroids in the outer disk.
It is not currently possible to directly measure $\alpha(r)$ in protoplanetary disks, although some observations
may constrain this parameter.
High-resolution imaging of disk regions by Pinte et al.\ (2016) and Dullemond et al.\ (2018) show that the sharpness of the gaps and rings  
in the outer portions of many protoplanetary disks requires low $\alpha < 10^{-4}$ more compatible with Cases I 
or III, but not with Case II.
Higher resolution imaging potentially could similarly constrain $\alpha$ in the inner disk in the future, but this
would be an observational challenge. 
Novel tests with sophisticated chemical models to detect heating from viscous accretion, such as detecting molecular 
emission in the cooler (90-400 K) layers of the disk, as put forth by Najita \& \'Ad\'amkovics (2017), or detect distinct 
chemical imprints that betray either a spatial variation in turbulence or a difference between low- and high-viscosity
disks  may also constrain $\alpha(r)$.

It would advantageous to use observations of water vapor in the inner portions of disks to constrain $\alpha$ there.
Differences in water vapor abundance take time to develop, but by 5 Myr, in our Case I uniform-$\alpha$ disk, 
with $\alpha = 3 \times 10^{-4}$ in the inner disk, the water vapor mass fraction is $1 \times 10^{-4}$.
For case II, $\alpha = 1 \times 10^{-3}$ in the inner disk and the water vapor mass fraction is $4 \times 10^{-4}$.
And in Case III, $\alpha = 1 \times 10^{-3}$ throughout a broader region in the inner disk, and the water
vapor abundance is $2 \times 10^{-4}$.
The relationship between $\alpha$ and water vapor abundance is therefore complicated, but forward models with
$\alpha(r)$ as an input should be able to test models against observations. 
A particular feature of our Case II disk that might be observable is the dip in water abundance by a factor of almost 2
just inside the snow line, between 2 and 3 AU. 

We highlight that the $\Sigma$ profile itself as well as the water abundance  $N_{\rm H_2O}/N_{\rm H_2}$ across the inner disk are by themselves a diagnostic of the $\alpha$ profile in the inner disk and thus the dominant mechanism of angular momentum transport operating in disks.

\subsubsection{Solar System Observations}

Our results suggest it may be possible to use the spatial distribution of asteroid water content to constrain 
the $\alpha(r)$ profile of the solar nebula disk.
The water content of asteroids is discernible using spectroscopy and comparison to meteorites.

Gradie \& Tedesco (1982) determined that different classes of asteroids are predominantly present in specific locations 
of the asteroid belt; S-type asteroids dominate in mass and number in the inner belt inside about 2.7 AU, while C-type
asteroids dominate in mass and number in the outer belt beyond about 2.7 AU (see Fig.\ 3 of DeMeo \& Carry 2014; the 
largest asteroids correspond to all the mass above the horizontal dotted line at $\sim 1 \times 10^{18}$ kg). 

S-type asteroids show little evidence of hydration features in their spectra, and are spectrally associated with
ordinary chondrites, which are generally water-poor, with only $0.1 - 1$ wt\% water (Hutchison et al.\ 1987; Alexander et al.\ 1989, 2013).
C-type asteroids do show hydration bands in their spectra, and are associated with water-rich carbonaceous chondrites, with up to $\sim 10$wt$\%$ (structurally bound) water (Alexander et al.\ 2013).
Additionally, E-type asteroids at 2.0 AU are spectrally associated with enstatite chondrites, which accreted essentially no water (Jacquet et al.\ 2017). 
Rarer R chondrites appear to have been water-rich and formed perhaps at 2.6 AU (Desch et al.\ 2018). 
DeMeo \& Carry (2014) were able to extend the study of Gradie \& Tedesco (1982) to include the wealth of information attained since then for the smaller asteroids down to 5 km in size, finding that while a sharp radial gradient exists for the distributions of the largest asteroids' spectral types (consistent with the conclusions of Gradie \& Tedesco 1982), the smallest asteroids of different classes have been significantly radially mixed. This is commonly interpreted as a radial mixing of asteroids scattered into that region, with S-types at r $<$ 2 AU scattering out, mixing with C-types from r $>$ 3 AU scattering in (e.g., the Grand Tack model of Walsh et al.\ 2011). This is a very viable interpretation of the data, but our modeling offers an alternative explanation, that asteroids in the 2-3 AU region may have formed in place, with dry asteroids forming early and wet asteroids forming later, after the snow line swept inward.

We imagine that asteroids form by a rapid mechanism like streaming instability, and are snapshots of the disk composition at that place and time. The number of asteroids created per time, per area of the disk, would be $\dot{N}$ = ($\Sigma$ / $t_{\rm growth}$) / $M$, where $\Sigma$ is the surface density of solids, and $M$ is the mass of a typical asteroid. For large radii that are always outside the snow line, all of the asteroids are ``wet". For small radii that are always inside the snow line, all of the asteroids are ``dry". For intermediate radii that the snow line sweeps through, some asteroids are wet and some are dry. Therefore at the time the asteroids in the inner disk formed, at $\approx 2 - 3$ Myr (Desch et al.\ 2018), the snow line must not have extended inward of 2.0 AU; and the asteroids formed between 2 AU and 3 AU or more around this time appear to have sampled a variety of water-ice fractions. 


Comparing these findings about the asteroid belt to our various disk models, Case II seems least consistent with our
asteroid belt. Its snow line does not move inward of 3.5 AU by 3 Myr, and the range of radii over which the asteroid 
water-ice fractions vary is narrow, $< 1$ AU. Case II is also marked by a region in which the water-ice fraction of
asteroids exceeds the canonical ratio. 
Case I is broadly consistent with the asteroid belt, with the snow line at about 2.5 AU at 3 Myr, and a broad range
of radii over which the asteroid water-ice fractions vary. The gradient of water-ice fractions would be monotonically
increasing with distance so that no asteroids would have ice fractions $> 50$\%, for $\alpha > 10^{-3}$.
Finally, Case III appears most consistent with the asteroid belt, with the snow line at 2.0 AU at 3 Myr, and a
monotonically increasing water-ice fraction in the inner disk.

None of these cases includes the effects of Jupiter opening a gap in the disk, which Morbidelli et al.\ (2016) have
shown to be potentially very important in setting the water content of the inner disk. 
Nevertheless, our analysis shows how one might use the distribution of asteroids and water content to infer 
$\alpha(r)$. Future work is encouraged, but so far disk evolution by MRI does not appear consistent with the constraints from 
our asteroid belt. 

We also note that different initial surface density profiles would lead to different amount of ice that would be locked beyond the snow line in chondrules and eventually asteroidal material. A shallower surface density profile would lead to more ice being accumulated beyond the snowline, than a more steeper profile. This is an intriguing result that warrants a more detailed parameter study.

\section{Summary and Conclusions}

\subsection{Summary} 

The distribution of water and other volatiles in planets depends on how these volatiles are transported through
the protoplanetary disk, as diffusing and advected vapor, and as diffusing, advected and drifting icy particles.
The partitioning of a volatile between vapor and ice depends on the temperature, and the ``snow line" in a disk, 
demarcating ice-rich and vapor-rich regions, is sensitive to the degree of accretional heating in a disk.
The transport of particles and diffusion of vapor and particles depends on the strength of turbulence transporting
angular momentum. 
These processes ultimately depend on the magnitude and spatial variation of $\alpha$, the turbulence parameter.

We have investigated the effects of different $\alpha(r)$ profiles on the distribution of water in protoplanetary disks.
We have considered three $\alpha(r)$ profiles in particular. 
Our Case I assumes a uniform value of $\alpha$ throughout the disk; our canonical case considers 
$\alpha = 3 \times 10^{-4}$, but we also have explored higher and lower values.
In Case II we investigate a profile of $\alpha(r)$ motivated by simulations of how the MRI would operate in protoplanetary
disks, with higher values of $\alpha$ close to the star, where temperatures are hot, and far from the star, where
low densities permit high ionization fractions, but low at intermediate radii, where the disk is dominated by 
MRI ``dead zones". 
This $\alpha(r)$ profile varies between $10^{-4}$ and $10^{-3}$, and is on average close to $3 \times 10^{-4}$. 
This profile is an approximation of the results of Kalyaan et al.\ (2015). 
In Case III, we adopt an $\alpha(r)$ profile similar to that of Desch et al.\ (2018) constructed to explain a variety of 
meteoritic data.
This profile is marked by low $\alpha = 1 \times 10^{-4}$ throughout the disk, but rising at intermediate radii to a 
an elevated value $\alpha = 1 \times 10^{-3}$ in the inner disk.  As in Case II, this profile varies between $10^{-4}$
and $10^{-3}$ and is on average close to $\alpha = 3 \times 10^{-4}$. 
This profile is consistent with hydrodynamic instabilities like vertical shear instability acting throughout the disk, 
augmented by magnetic disk winds in the inner disk.

With these $\alpha(r)$ profiles as inputs, we conducted 1-D disk simulations including the transport of vapor and small
particles by advection and diffusion, plus transport of chondrule-sized (1 mm) particles by advection, diffusion, and 
radial drift due to aerodynamic drag. 
We included the condensation of vapor to ice on chondrule surfaces and the sublimation of ice to vapor.
We also accounted for growth of planetesimals from the population of chondrules or icy chondrules, predicting the 
ice fractions of these asteroidal bodies formed at each distance from the star. 

\subsection{Important Conclusions}

We present the following important results from this study:
\begin{enumerate}
\item Radial water distribution in the protoplanetary disk is sensitive to the subtly-interconnected roles of the various physical processes of advection, diffusion and drift of vapor and solids. However, factors that affect drift play the most important role in setting abundances of vapor and ice inside and outside of the snow line, in disk gas and solids, and eventually planetesimals. The rates of drift determine how much ice mass is carried into the inner nebula, half of which is inevitably moved outward via vapor diffusion through the snow line, leading to high water-rock ratios for chondrules here. If chondrule diffusion is efficient beyond the snow line, these icy chondrules are less likely to drift back towards the snow line, and are accreted into asteroids that remain there.
\item Disks with lower (global) $\alpha$ show higher peak abundances of water in asteroids right beyond the snow line, with a narrower peak width. This translates to sharper volatile gradients in asteroids that may form at different distances. Slight enhancements and depletions in the radial abundance profile are maintained for a longer duration in less viscous disks, which may provide unique localized environments for interesting chemistry. Vapor abundance in the inner disk also stays relatively high. 
\item Disks with higher $\alpha$ lead to efficient depletion of water vapor in the inner disk over few Myr timescales. Due to enhanced diffusivity of chondrules beyond the snow line, a shallower peak abundance with a broad peak width results, and the snow line sweeps a greater distance with time. Such disks evolve more rapidly, leaving less time for planets to accrete mass, unlike less viscid disks and even radially varying $\alpha$ disks that have less turbulent regions.
\item While the early location of the snow line is dependent on $\alpha$, the final location (at 5 Myr) is not sensitive to the choice of the globally uniform $\alpha$.
\item Ultimately, for uniform $\alpha$ disks, the order of importance of the various factors determining the magnitude of the peak ice fraction in chondrules beyond the snow line is as follows: $a_{\rm chon}\, > \, t_{\rm growth}\,>\,\alpha > Sc > \kappa $, i.e., planetary properties are more important than disk properties. The order of importance of factors affecting the peak ice fraction in asteroids beyond the snow line is as follows: $\alpha > t_{\rm growth} > a_{\rm chon} > $ everything else. Here, disk and planetary parameters seem to be as important as the other.
\item The inner disk water vapor content is sensitive to the size of the drifting solids that brings water inward from the outer nebula. Even a slight increase in particle size by a factor of 2 (i.e., $a_{\rm chon}$ = 0.03 to 0.06 cm) produces a persistent enhancement in water vapor in the inner disk interior to the snow line for at least $\sim$ 0.5 Myr.
\item Assuming a more rapid growth timescale of asteroids (i.e., 0.3 Myr; comparable to that of chondrule and vapor diffusion) results in the depletion of the chondrule pool beyond the snow line before they diffus inward, as well as back diffusion of the vapor inward of the snow line diffusing outward, thus affecting the subtle `equilibrium' achieved by these diffusive processes. This causes the snow line itself to slightly move outward.
\item Radially varying $\alpha$ disks yield disk structure distinctly different from the smooth uniform $\alpha$ disk. An MRI-active disk may have a tenuous inner disk with a mass pile up at distances of a few AU. On the contrary, the hybrid $\alpha$ case with the more turbulent inner disk may have a flat (and low) $\Sigma(r)$ upto 20 AU within 0.5 Myr. These effects consequentially lead to different radial distribution of volatiles in each disk.
\item In CASE II (MRI-$\alpha$ disk), much of the bulk gas (along with the vapor and chondrules) collects at around 1 AU. The snow line also does not move much with time. Both of these processes have the effect of producing large ice-rock ratios of chondrules beyond the snow line. 
\item For the hybrid $\alpha$ disk, higher $\alpha$ in the inner disk leads to high accretion rates, quick depletion of the disk mass, and therefore drastic drop in the temperature of the inner disk. This leads to significant movement of the water snow line that does not allow for abundances of ice to reach locally as high values as in CASE II. 
\item We argue that CASE III appears most like the solar nebula, and yields asteroids with a diverse population of differing water content as well as a monotonically increasing amount of water content. This scenario is also consistent with the formation of enstatite chondrites at $<$ 2 AU, ordinary chondrites at 2-4 AU region (after a few Myr) and carbonaceous chondrites beyond 3-4 AU (Desch et al.\ 2018). This suggests that we might be able to constrain $\alpha(r)$ from the radial distribution of asteroidal water content.

\item In summary, we argue that $\Sigma(r)$, radial water abundance  $N_{\rm H_2O}/N_{\rm H_2}$ as well as radial distribution of asteroidal water content provide additional avenues for constraining $\alpha(r)$ and therefore the mechanism of disk evolution.
\end{enumerate}

\acknowledgments

The authors are grateful for helpful discussions with Jacob Simon, Kevin Flaherty, Cornelis Dullemond and Prajkta Mane. The authors are also grateful for the insightful suggestions from the anonymous reviewer.




\end{document}